\documentclass[12pt]{iopart}
\usepackage{amssymb}
\usepackage{txfonts}

\usepackage{subfigure}
\usepackage[pdftex]{graphicx}
\usepackage{booktabs}
\usepackage{eurosym}
\usepackage{epstopdf}

\usepackage{ulem}
\usepackage{color}

\begin{document}

\title{Comparison of bidirectional pedestrian flows by experiments}
\author{J Zhang$^1$ and A Seyfried$^{1,2}$}

\address{$^1$ J\"ulich Supercomputing Centre, Forschungszentrum
 J\"ulich GmbH, 52425 J\"ulich, Germany}
\address{$^2$ Computer Simulation for Fire Safety and Pedestrian Traffic,
 Bergische Universit\"at Wuppertal, Pauluskirchstrasse 11, 42285 Wuppertal, Germany}
\ead{ju.zhang@fz-juelich.de, seyfried@fz-juelich.de}

\begin{abstract}



Intersections of pedestrian flows feature multiple types, varying in the numbers of flow directions as well as intersecting angles. In this article results from diverse bidirectional flow experiments with different intersecting angles are compared. To analyze the transport capabilities the Voronoi method is used to resolve the fine structure of the resulting velocity-density relations and spatial dependence of the measurements. The fundamental diagrams of various flow types are compared and show no apparent difference with respect to the intersecting angle. This result indicates that head-on conflicts of different types of flow have the same influence on the transport properties of the system, which demonstrates the high self-organization capabilities of pedestrians.

\end{abstract}

\section{Introduction}




Diverse bidirectional pedestrian flows both with and without intersection can be observed in places like crosswalks, sidewalks as well as stairways et al..  In contrary to vehicular traffic, these streams are usually not regulated by traffic lanes, rules or lights. Thus conflicts among pedestrians in bidirectional flow can be more intense. In certain circumstance it could lead to serious consequences like Love Parade disaster in Duisburg 2010 \cite{duisburg}, where multidirectional pedestrian streams led to a congestion at the entrance to a music event.

Bidirectional flows could be influenced by series of factors including the intersecting angles ($\varphi$), the directional flow ratio ($r$), the boundary conditions, the width of the stream or the facility($b$) as well as human factors amongst others. In this paper we define the intersecting angle $\varphi=0^{\circ}$ if both streams flow in the same direction and $\varphi=180^{\circ}$ if the stream directions are exactly opposite. In \cite{Zhang2012} we collect fundamental diagrams of bidirectional flow from different empirical studies. Most of the data were measured at low density situation for $\rho < 2.0~m^{-2}$ and show large discrepancies. Since these data are obtained under different experimental situations and different measurement methods, it is difficult to identify the origin of these differences. In the following we discuss this in more detail.

First we focus on the question whether the directional flow ratio has influence on the capacity. In Highway Capacity Manual (2000) it is mentioned that for bi-directional pedestrian streams of roughly equal flow in each direction, only a slight reduction of the capacity occurs. The manual suggests that the maximal reduction in the capacity sets in at a directional split ratio of 0.9 versus 0.1 \cite{HCM2000}. Lam et al. also found that the maximum reduction in capacity is around 19\% and it happens when directional split ratio is 0.9 versus 0.1 \cite{Lam2002, Lam2003}. On the contrary, Wael et al.  found that maximal reduction in the capacity occurs when the directional split ratio is 0.5 at signalized crosswalks \cite{Alhajyaseen2011}. They assume that this variation could be traced back to different behavior at long walkways and short signalized crosswalks. However, different definitions and measurements of the flow ratio are used in these studies and makes a comparison impossible. In contrary to these studies, we don't find a reduction of the flow for different ratios when the density is smaller than $2~m^{-2}$ \cite{Zhang2012}.

Second we discuss the influence of the intersecting angle. Existing empirical studies mainly focus on bidirectional pedestrian flow at crosswalks \cite{Navin1969, O'Flaherty1972, Polus1983,Tanaboriboon1986, Lam2002, Jian2010} and shopping streets \cite{Older1968} where the opposing flows have an intersecting angle close to $180^{\circ}$. In \cite{Hughes2000,Hoogendoorn2003153, Cividini2013} crossing pedestrian flows are studied numerically and formation of pattern are discussed only qualitatively without considering the influence of intersection angle. Wong et al. conducted controlled experiments in which two groups of students were asked to walk on designated walkways with different levels of pedestrian flow different intersecting angles \cite{Wong2010, Guo2012c}. The experiment was mainly used to obtain data for the development of bidirectional pedestrian stream model but was unfortunatly not analyzed in detail due to technical restriction on the extraction of trajectories. In \cite{Zhang2012} we found differences regarding stability of the lanes formed in bidirectional flows between small intersecting angles and head on streams. For the former a clear separation of lanes occurs which is stable, while for the latter dynamical formation and extinsion of lanes are observable. However the fundamental diagrams show no difference in the density ranges observed.

Next we discus the impact of conflicts in bidirectional streams. The fundamental diagram of uni- and bidirectional flow \cite{Zhang2012} show obvious differences. We identified that conflicts of persons moving in the opposite direction reduce the speed of pedestrians leading the lane and reducing the speed of persons follwing in this lane. In bidirectional flow with nonzero intersecting angle, the area where the streams meet and conflicts occur is minimal when the intersecting angle is $90^\circ$. If the intersection angle rise from $90^\circ$ to $180^\circ$ the area where potentialy conflicts occur increases. This leads to the question whether reduction of flow in bidirectional streams depend from the angle of intersection.


The above discussion shows that up to now there is no consensus about the origin of the discrepancies between different types of bidirectional pedestrian flow.

The influences of boundary conditions, intersecting angle as well as the influence of conflicts on the fundamental diagrams are still not analyzed in detail. The aim of our study is to compare the fundamental diagrams of different bidirectional flows and analyze the influence of head-on conflicts on the transport properties. We will also study the effects of boundary conditions and geometries on the fundamental diagrams of bidirectional streams.

The remainder of the paper is organized as follows. In Section~\ref{sec-exp} we describe the setup of the experiments.  Section~\ref{sec-funda} describes the methods for data extraction and compares the fundamental diagrams for different types of streams. In last part we reveal the main conclusions in Section~\ref{sec-summary}.

\section{Experiments}\label{sec-exp}

In this paper, four different scenarios are considered to investigate the transport properties of various bidirectional pedestrian flow. All of them were peformed under laboratory conditions and realized different intersecting angles and boundary conditions. Scenario 1 was part of  the Hermes project and performed in the fairground D\"usseldorf in 2009 \cite{Holl2009, KemlohWagoum201398}. While the Scenario 2 to 4 were conducted by Plaue et al. in the entrance area of the Department of Mathematics building of Technische Universit\"at Berlin during the Long Night of the Science 2010 \cite{Plaue2011}.

\subsection{Scenario 1}

\begin{figure}
\centering\subfigure[SSL]{
\includegraphics[width=0.4\textwidth]{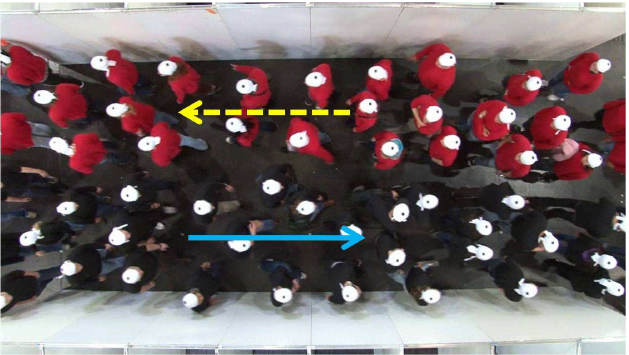}
\includegraphics[width =0.4\textwidth]{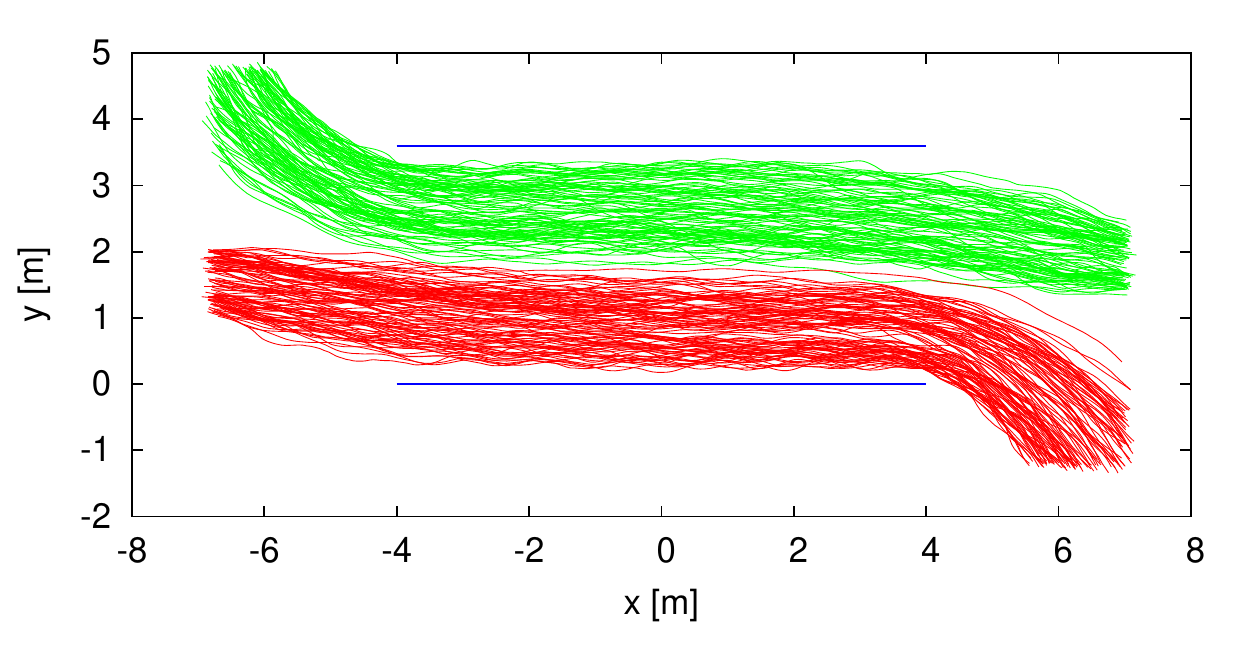}}
\centering\subfigure[DML]{
\includegraphics[width=0.4\textwidth]{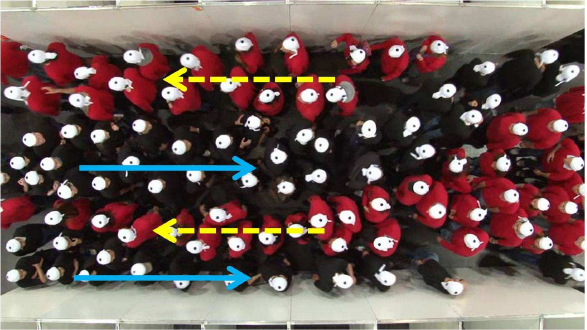}
\includegraphics[width =0.4\textwidth]{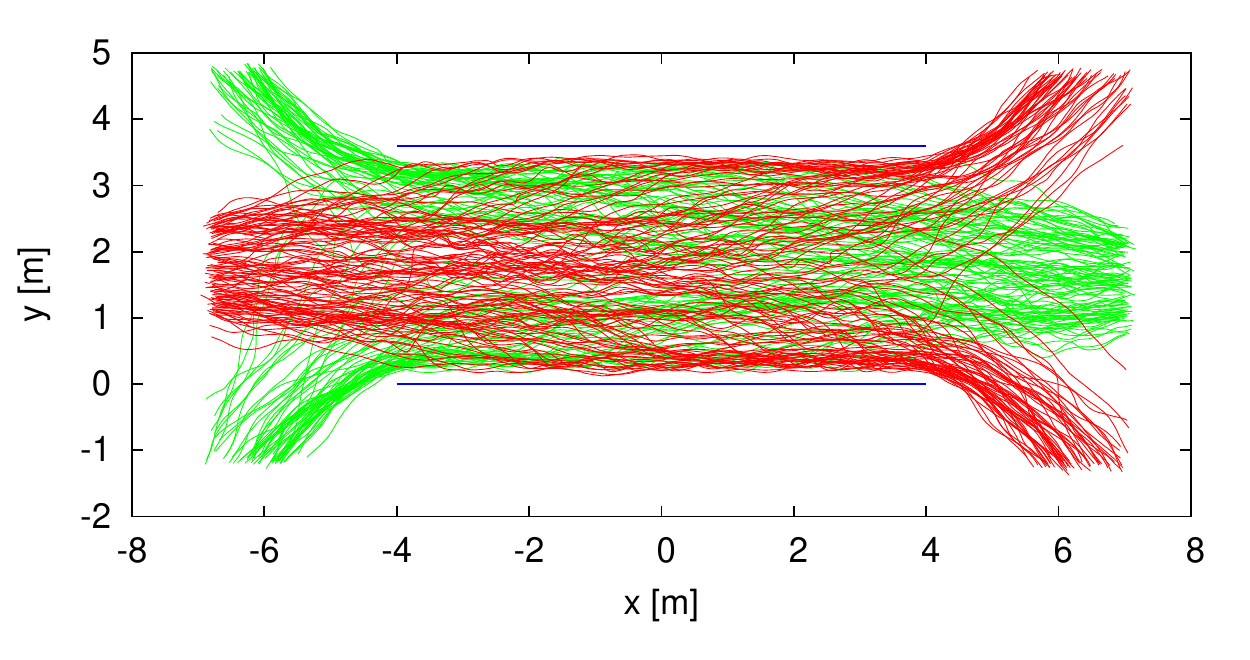}}
\caption{\label{hermes-exp} Snapshots and pedestrian trajectories from Scenario 1.}
\end{figure}

Figure~\ref{hermes-exp} shows snapshots and pedestrian trajectories of bidirectional pedestrian flow with separated single lanes (SSL) and dynamical
multi-lanes (DML). 22 runs were carried out in a $3.6~m$ wide corridor which was built on plane ground by boards.
In each run, the inflow rate into the corridor was changed by varying the entrance widths to control the density in the corridor. The participants were given
different instructions on exit selection to get different kinds of lanes. The details of this experiment have been described in \cite{Zhang2012}.

Intuitively, there should be more head-on conflicts in DML flow than in SSL flow, which maybe influence the fundamental diagram of them. However, no apparent difference is observed in our study at least for density $\rho < 2.0~m^{-2}$, which is the highest density obtained from our SSL flow experiment. Further, the specific flow concept is also applicable at the observed density ranges. Based on these considerations, we focus on the DML flow experiments in $3.6~m$ wide corridor, which has a wider range of density.

\subsection{Scenario 2}

\begin{figure}
\centering{
\includegraphics[width=0.4\textwidth]{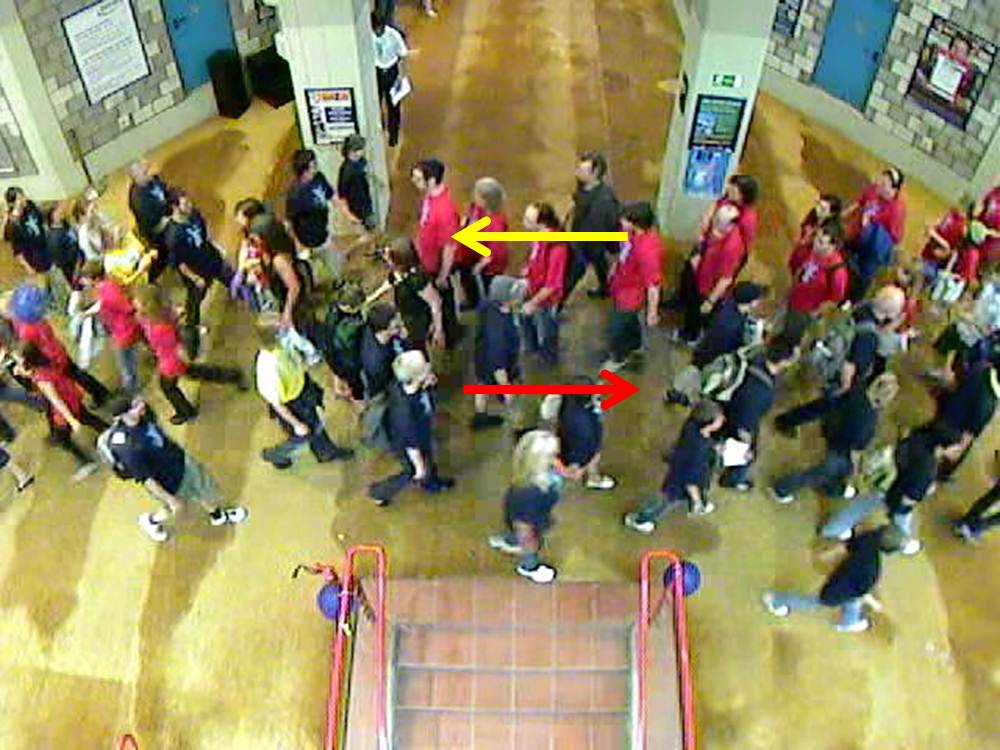}
\includegraphics[width =0.4\textwidth]{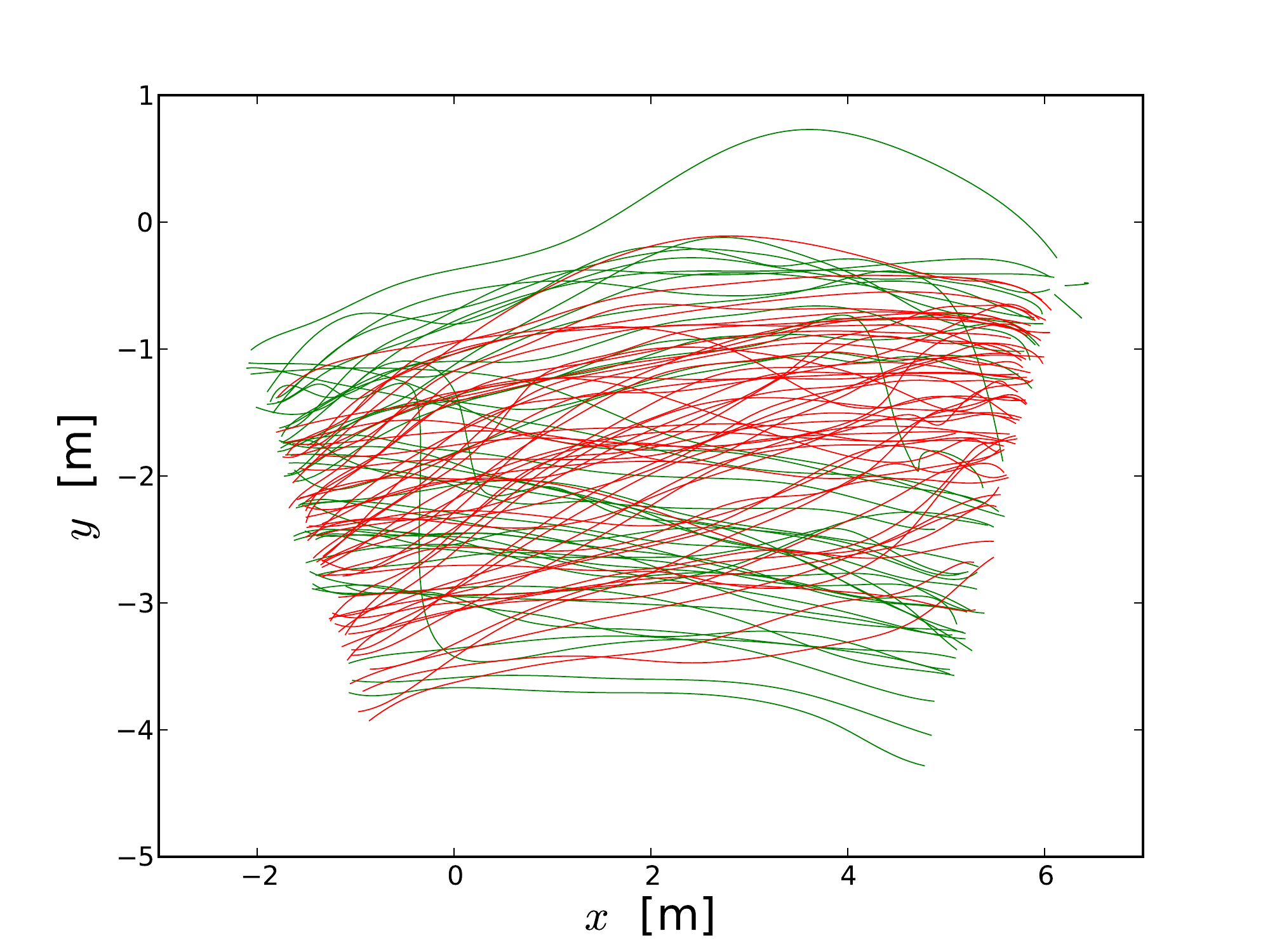}}
\caption{\label{br180-exp} Snapshots and pedestrian trajectories from Scenario 2.}
\end{figure}

In Scenario 2 two groups with 46 participants move on plane ground from opposite direction (see Figure~\ref{br180-exp}).
Unlike Scenario 1, the boundary of stream was given by the stream itself instead of walls in this scenario. Thus, pedestrians at the border of the stream have more space to move freely. Only one run was conducted with this scenario. In this process, three separated lanes and less head-on conflicts were observed due to the open boundary.

\subsection{Scenario 3}

\begin{figure}
\centering{
\includegraphics[width=0.4\textwidth]{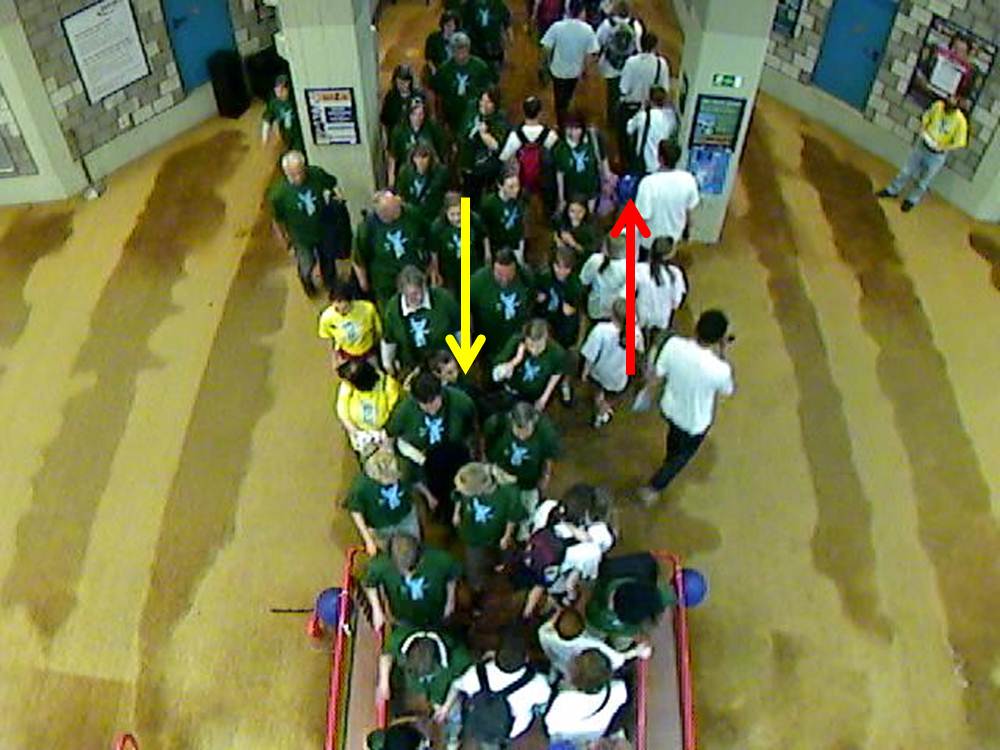}
\includegraphics[width=0.4\textwidth]{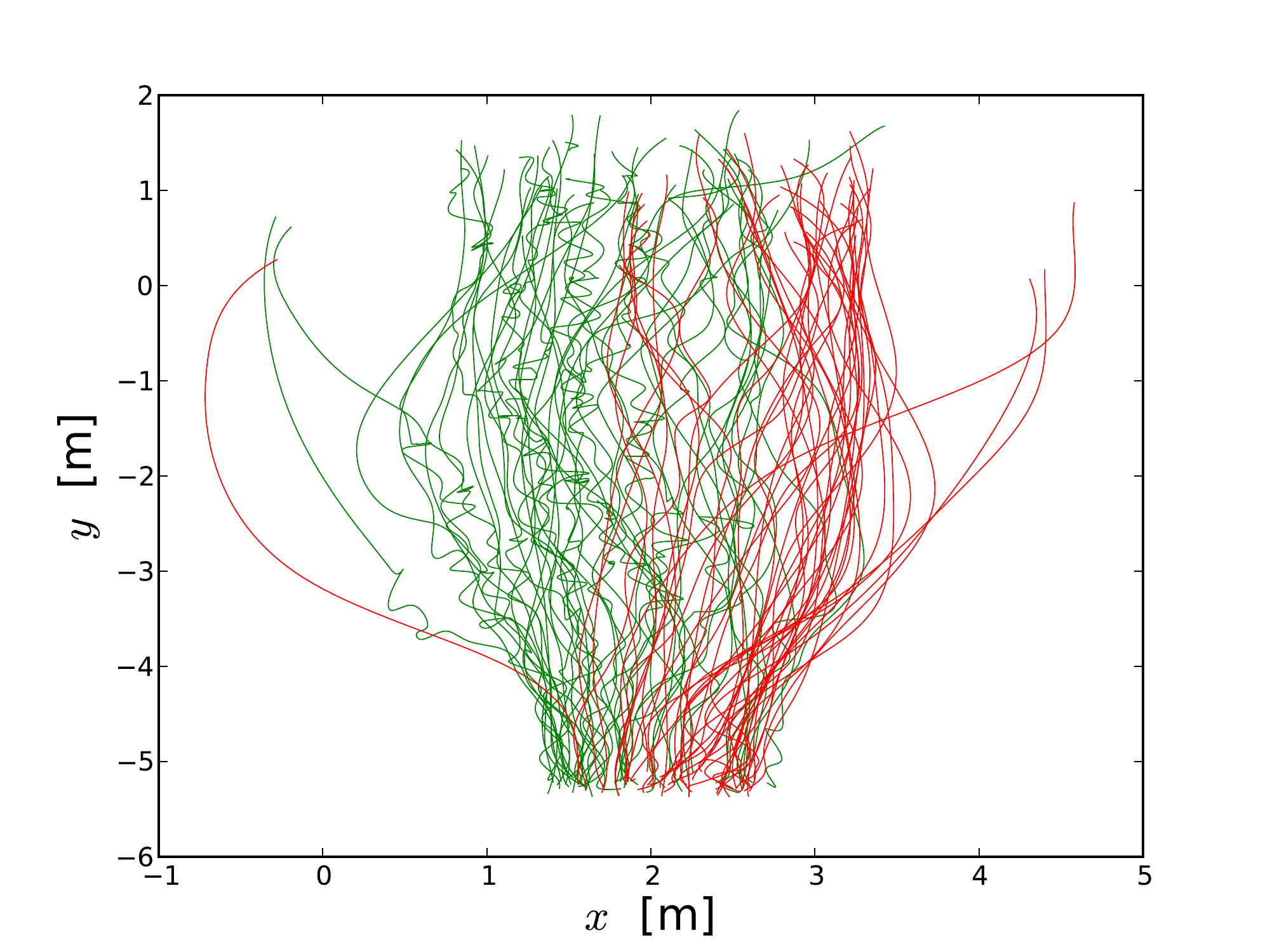}}
\caption{\label{gw180-exp} Snapshots and pedestrian trajectories from Scenario 3.}
\end{figure}

This scenario is also designed to get bidirectional flow with intersecting flows intersecting at $180^\circ$. Especially, a stairway
and two pillars were included in the scenario. That is to say, one group started from the staircase upward to the plane
ground and the other group moved from plane ground to staircase (see Figure~\ref{gw180-exp}). Here the two pillars
are used as potential boundaries on the plane ground and pedestrians should move in the area defined by them and stairway.
Actually, there were still few people not obeying the instruction. The stairway here was more like a bottleneck and head-on conflicts mainly occurred in the intersection.

\subsection{Scenario 4}

\begin{figure}
\centering{
\includegraphics[width=0.4\textwidth]{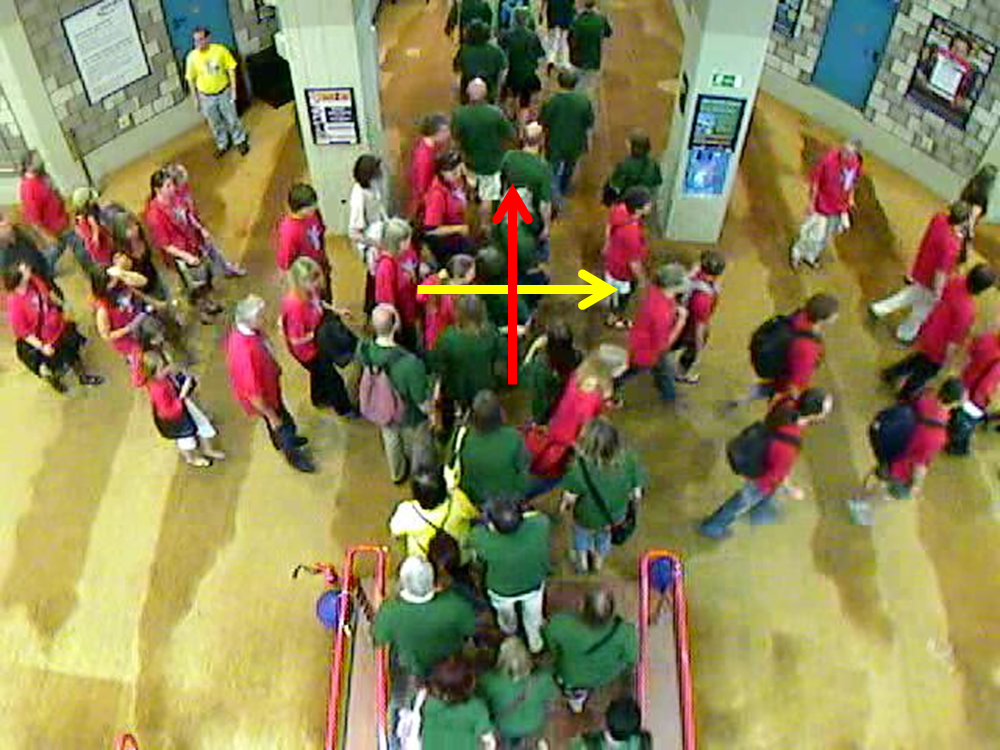}
\includegraphics[width=0.4\textwidth]{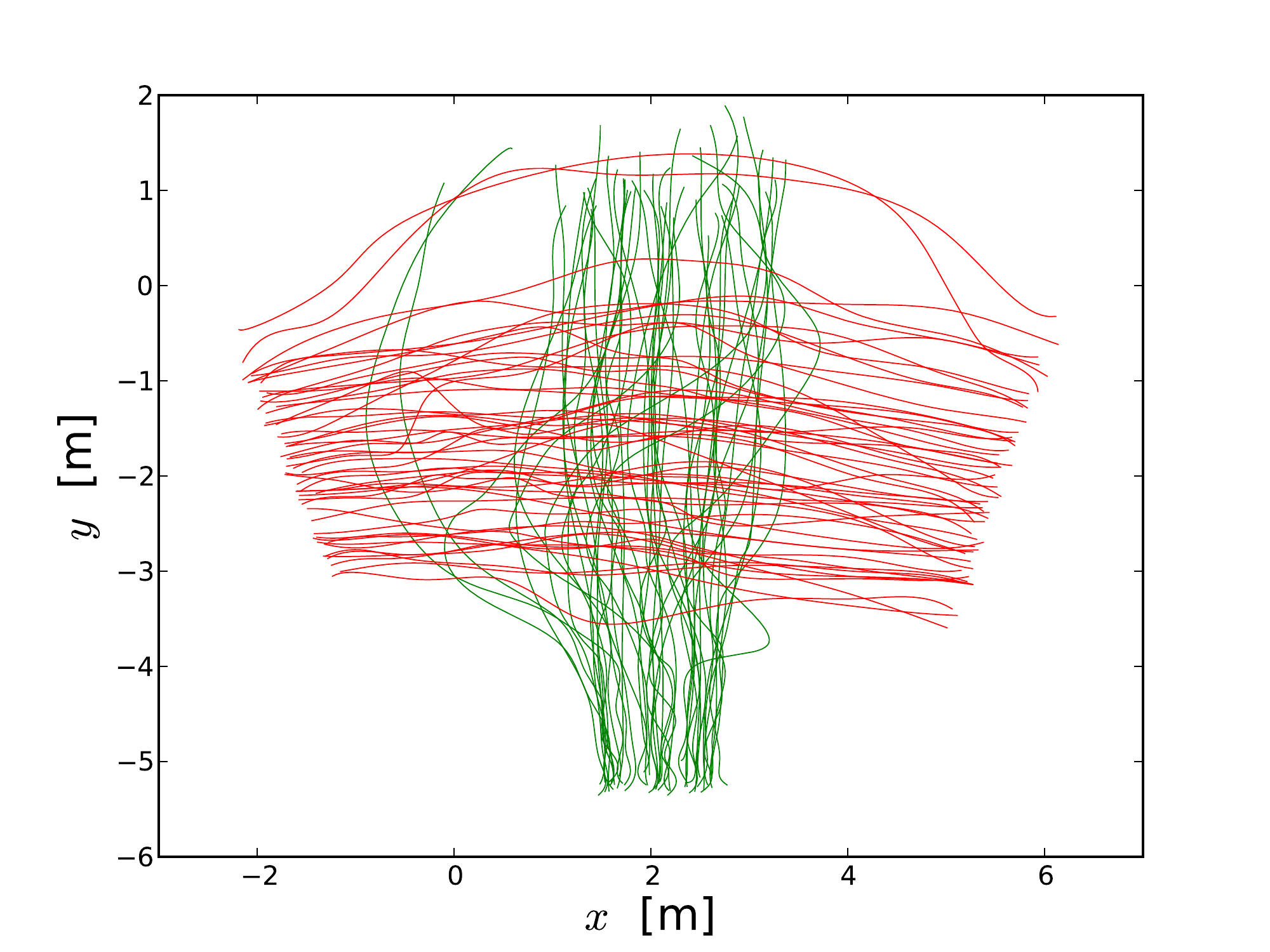}}
\caption{\label{gr90-exp} Snapshots and pedestrian trajectories from Scenario 4.}
\end{figure}

In this scenario, intersecting pedestrian flow from perpendicular directions was created. As shown in Figure~\ref{gr90-exp}, one group
moved from left to right on plane ground while the other group moved from ground to staircase perpendicularly. Facing the crossing flow in front,
pedestrians seemed to decelerate but not to cross vertically so as to avoid conflict with the ones appearing in front of them suddenly. This self-organized
coordination behavior plays an important role in this scenario.

\subsection{Comparison}

The participants in Scenario 1 were composed mostly of students with the mean age and height of $25 \pm 5.7~years~old$
and $1.76 \pm 0.09~m$ respectively. Whereas in Scenario 2, 3 and 4, they were volunteers from the event visitors and cover a wide range of
age including children. At the end of the event, they were instructed to move from opposite sides of the observation
area to create various scenarios. In both of these experiments, the participants were asked to move in normal speed.
The experiments were recorded by video cameras and pedestrian trajectories were extracted automatically from the videos.

\begin{figure}
\centering{
\includegraphics[width=0.9\textwidth]{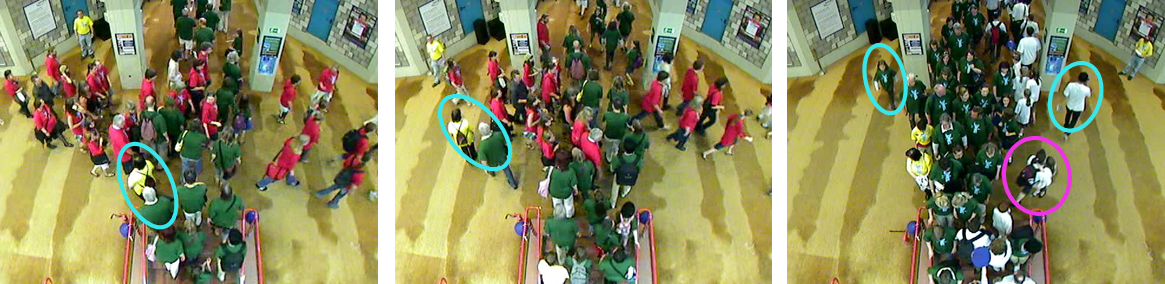}}
\caption{\label{setup-comp} Special pedestrian behaviors appeared in Scenario 2 to 4.}
\end{figure}

Compared to the experiments in the Hermes project, the experiments conducted in Berlin have larger free space. Firstly, the geometries in the latter experiment were open boundary without fixed walls due to constructional limitations. In this case, the pedestrians at boundary of stream could move freely and far way from the stream without following the instruction of the organizers (see blue circles in Figure~\ref{setup-comp}). Secondly, the participants of the latter experiment were visitors of a event but not specifically to the experiment. Some of them carried a bag during the movement, especially a participant was found taking a baby on his arm (see red circle in Figure~\ref{setup-comp}). Thirdly, in Hermes project series of runs were carried out for a certain scenario to obtain data at different density
ranges. However, in the latter experiment only one run was done for each scenario due to the time limitation and experimental conditions.
In this case we can only compare the results for part of density ranges in this study.

\section{Analysis and Results} \label{sec-funda}

\subsection{Measurement Method and Data Selection}

In this study, the Voronoi method is used to analyze the above mentioned experiments for its advantages of high resolution and small fluctuation of measured densities. Voronoi diagrams can be generated for every set of positions at time points given by the frame of the video recording. Based on these diagrams, the density and velocity distributions over space are obtained.  Then the Voronoi density and velocity for certain measurement area can be calculated further. For the details can be seen in \cite{Zhang2011}.

\begin{figure}
\centering{
\includegraphics[width=0.45\textwidth]{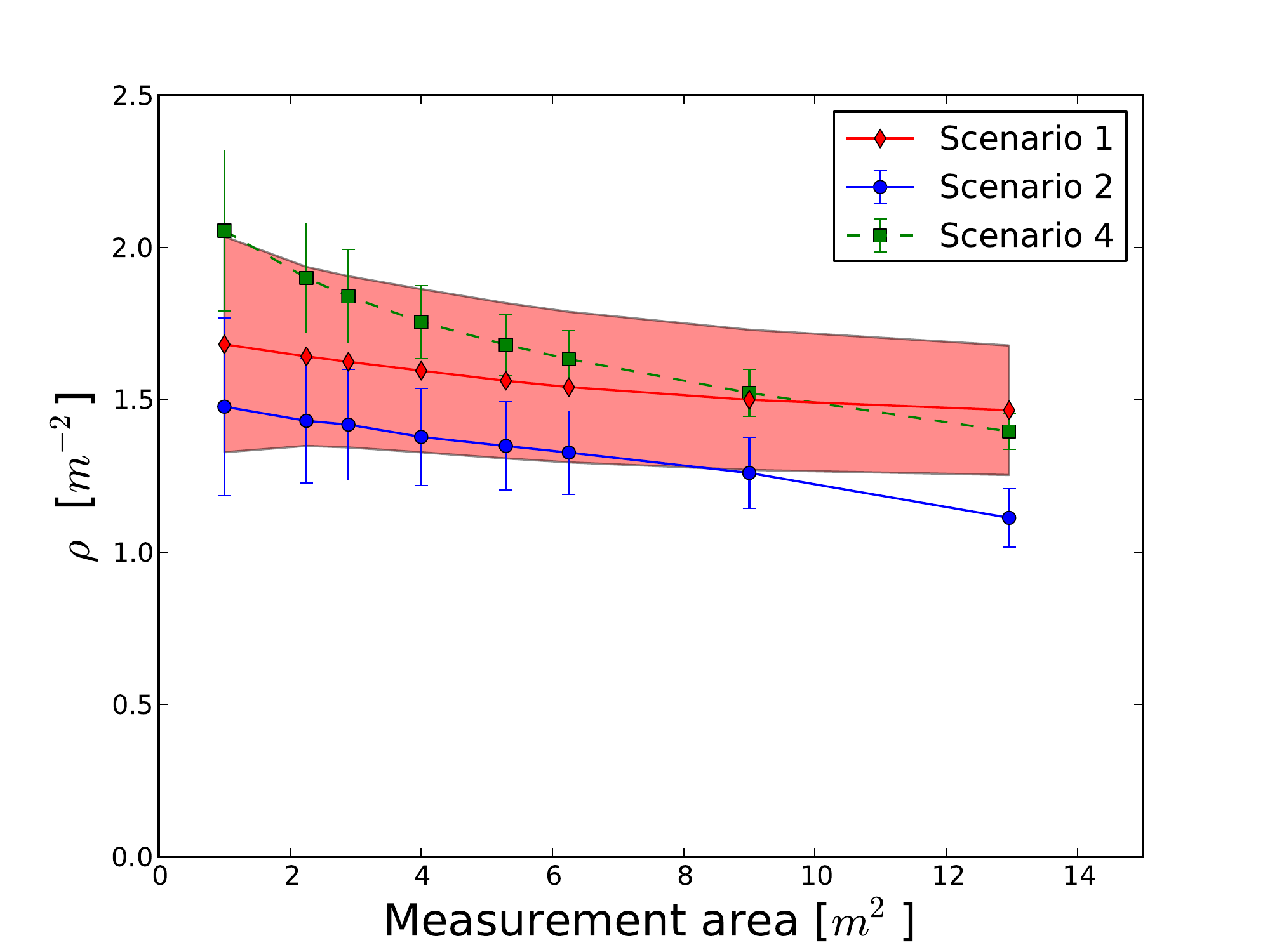}
\includegraphics[width=0.45\textwidth]{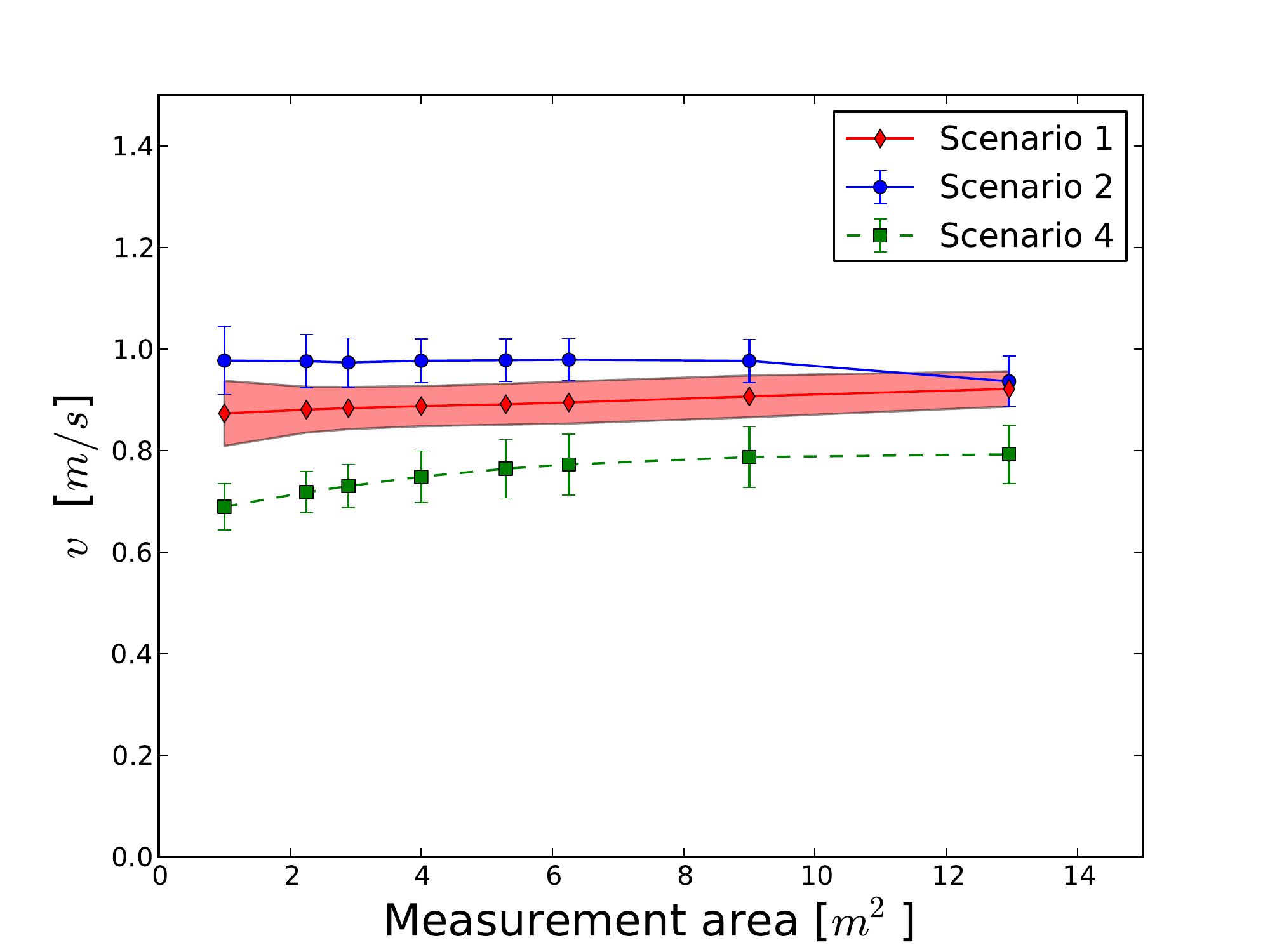}}
\caption{\label{fig-voro-measurearea} The influence of measurement area on the measured densities and velocities for one run in different scenarios. For Scenario 1, the standard deviations are presented by the ribbon.}
\end{figure}

\begin{figure}
\centering\subfigure[Scenario 1]{
\includegraphics[width=0.4\textwidth]{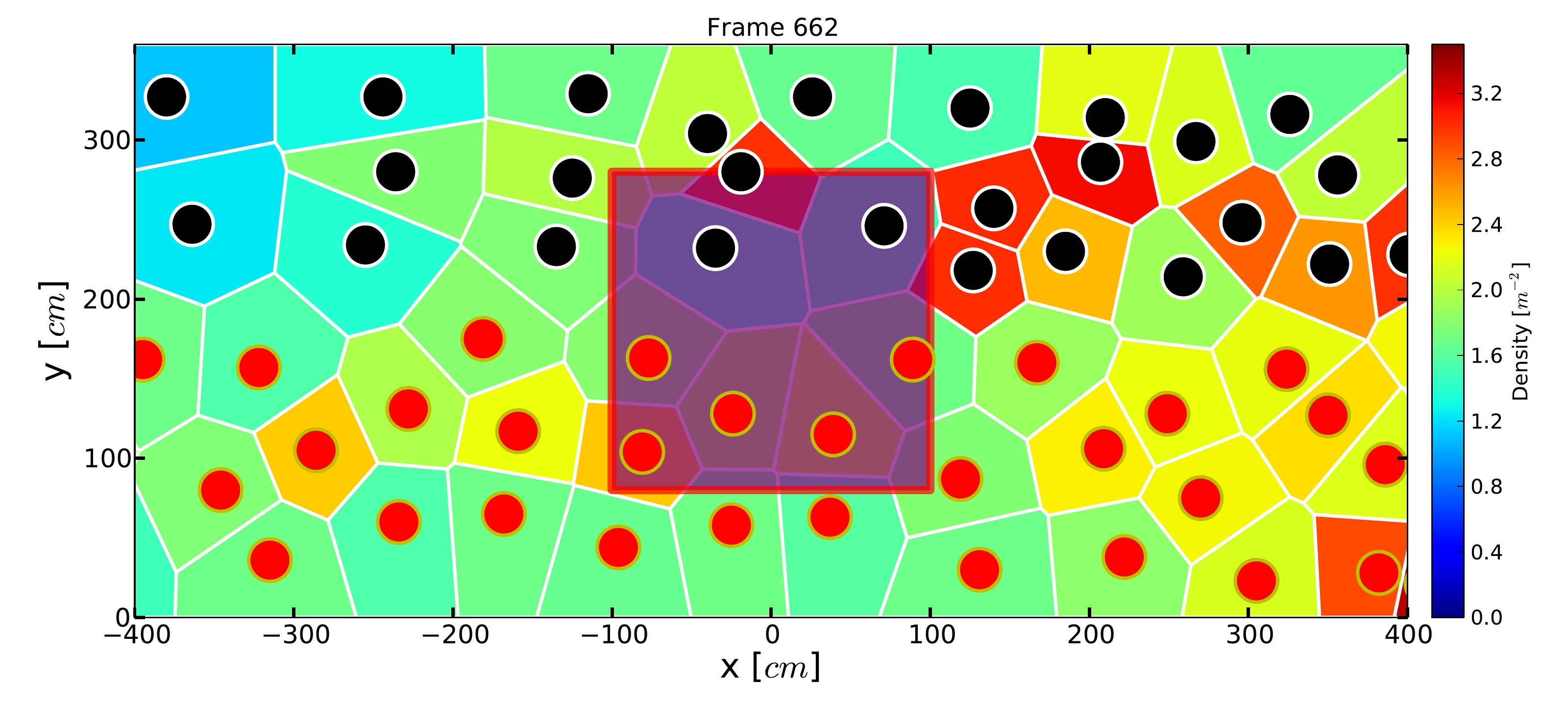}
\includegraphics[width=0.4\textwidth]{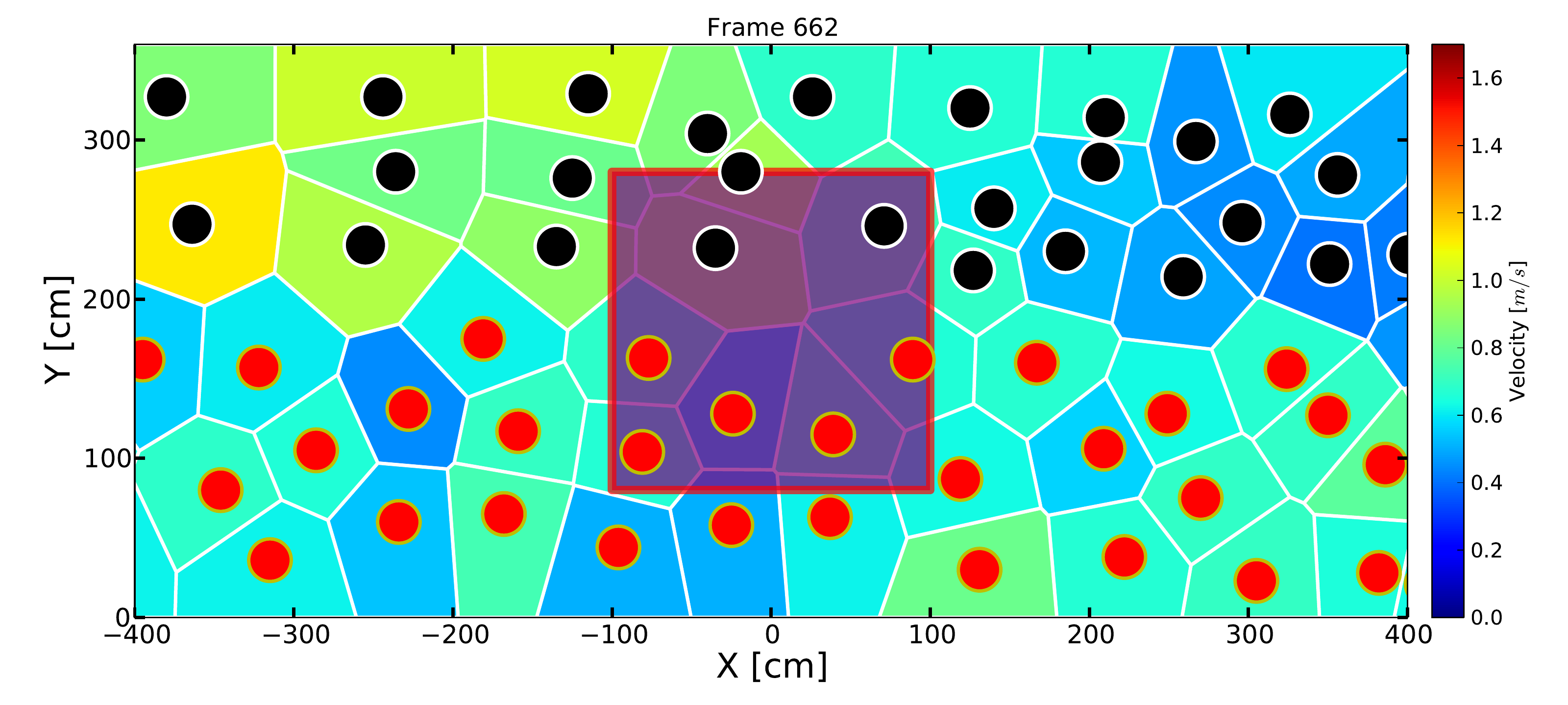}}
\centering\subfigure[Scenario 2]{
\includegraphics[width=0.4\textwidth]{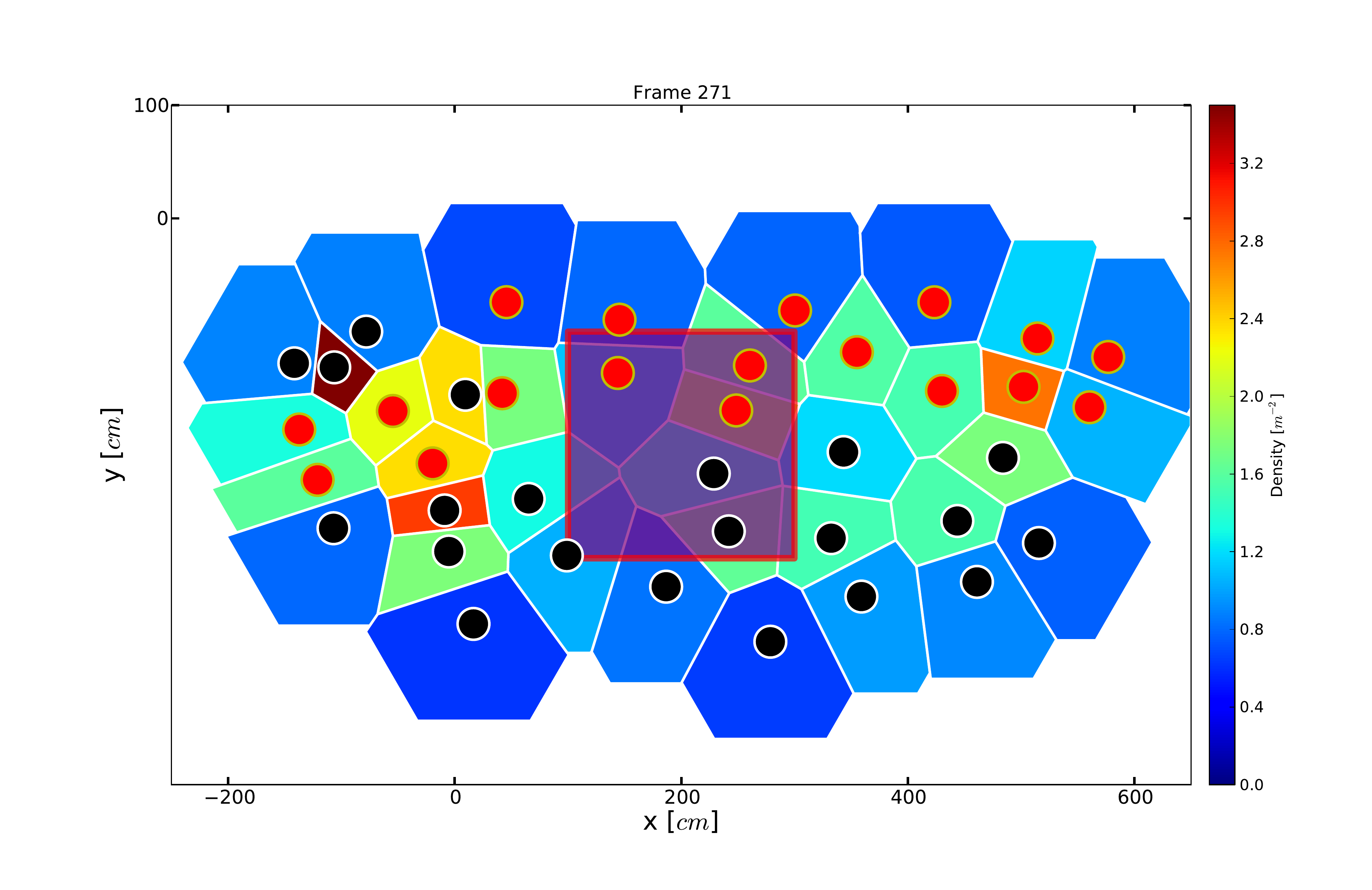}
\includegraphics[width=0.4\textwidth]{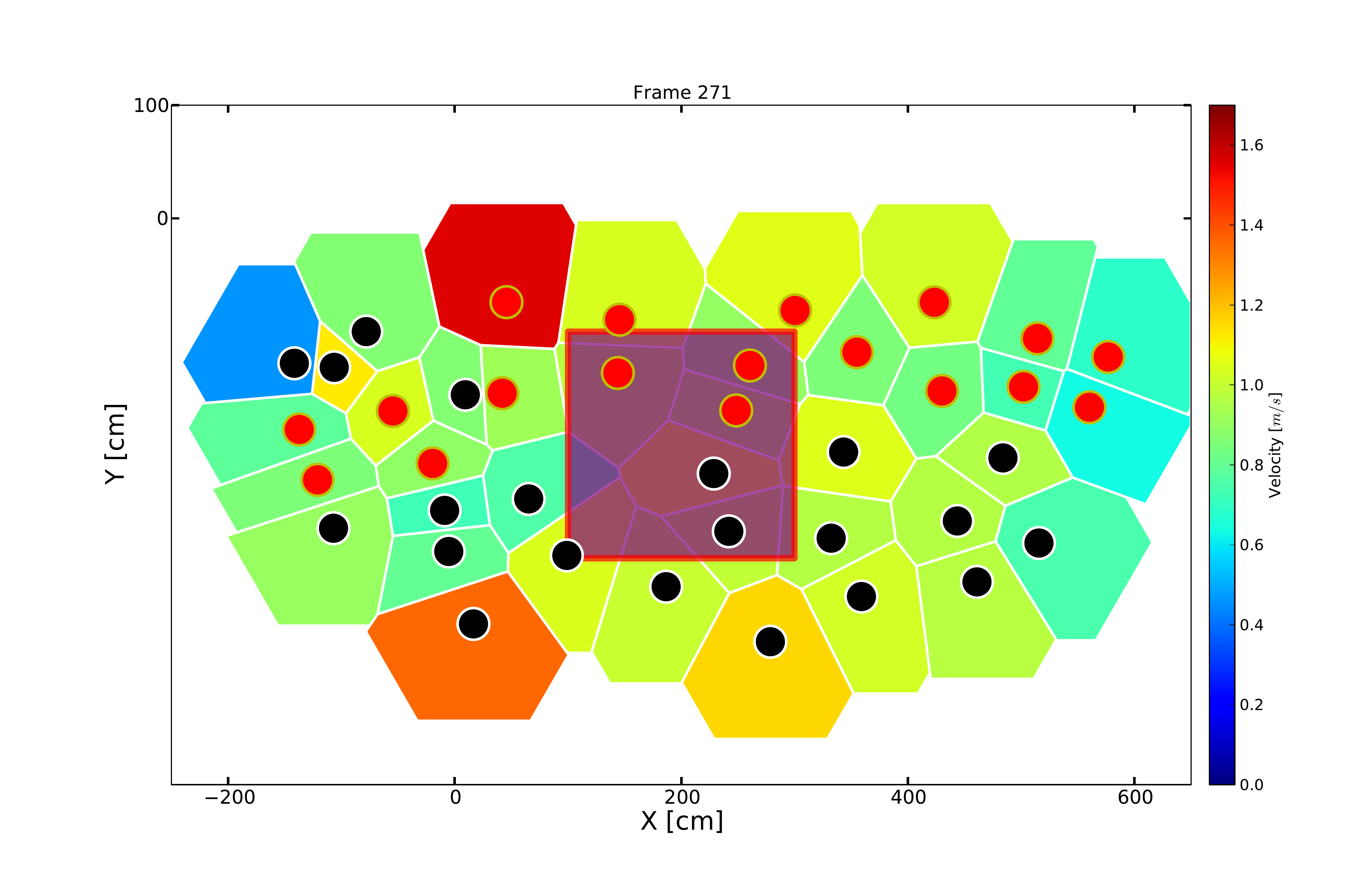}}
\centering\subfigure[Scenario 3]{
\includegraphics[width=0.4\textwidth]{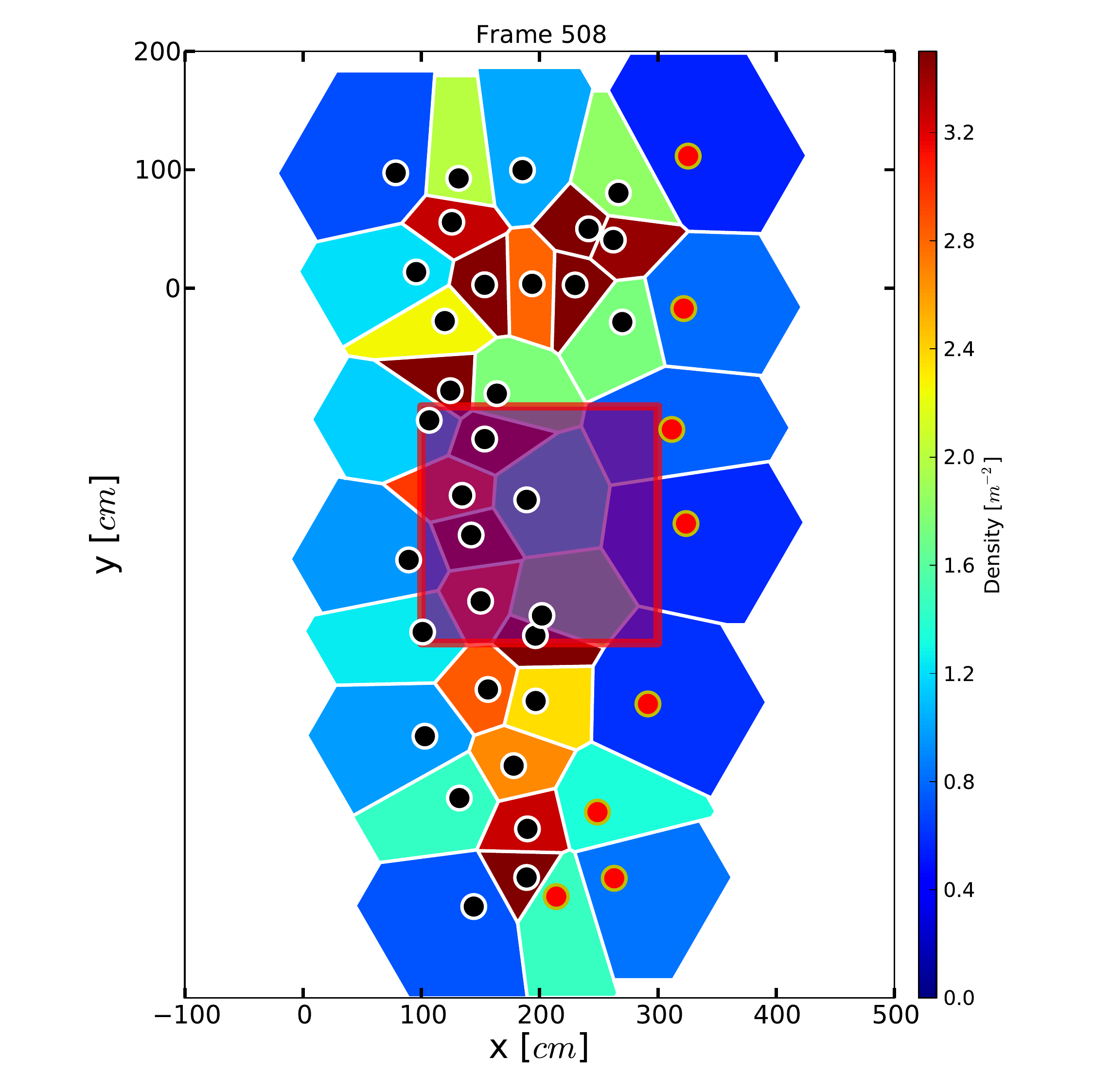}
\includegraphics[width=0.4\textwidth]{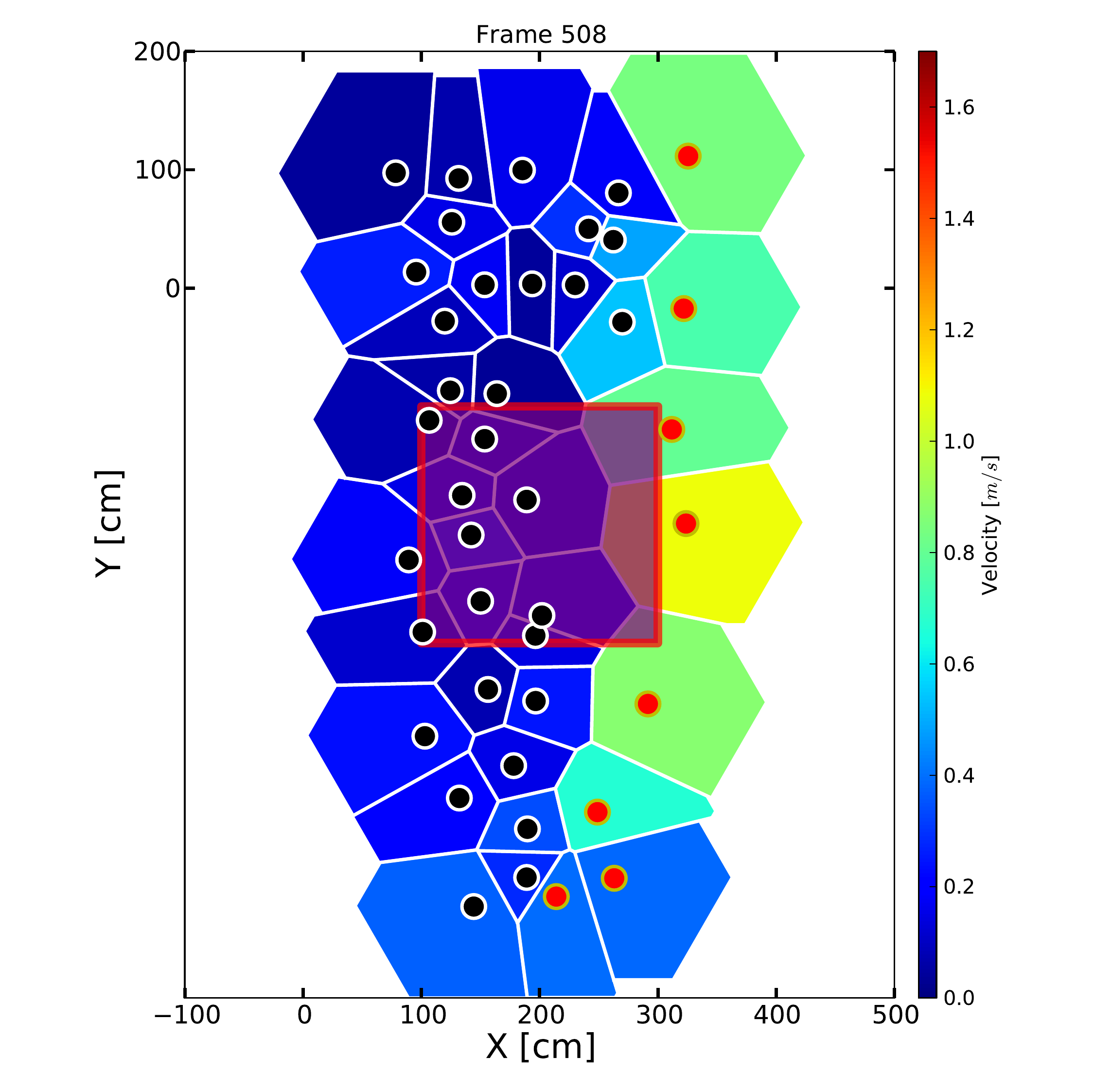}}
\centering\subfigure[Scenario 4]{
\includegraphics[width=0.4\textwidth]{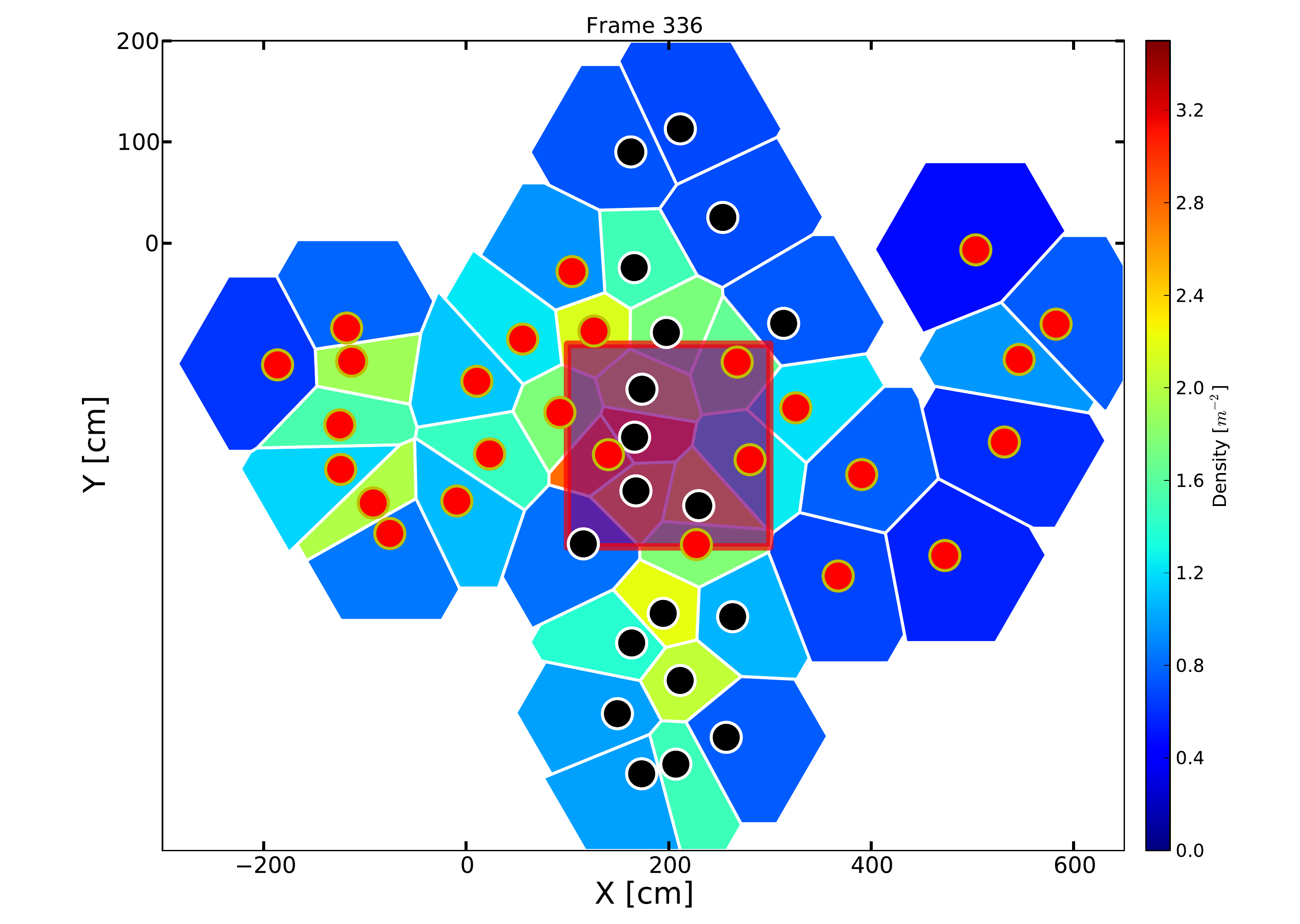}
\includegraphics[width=0.4\textwidth]{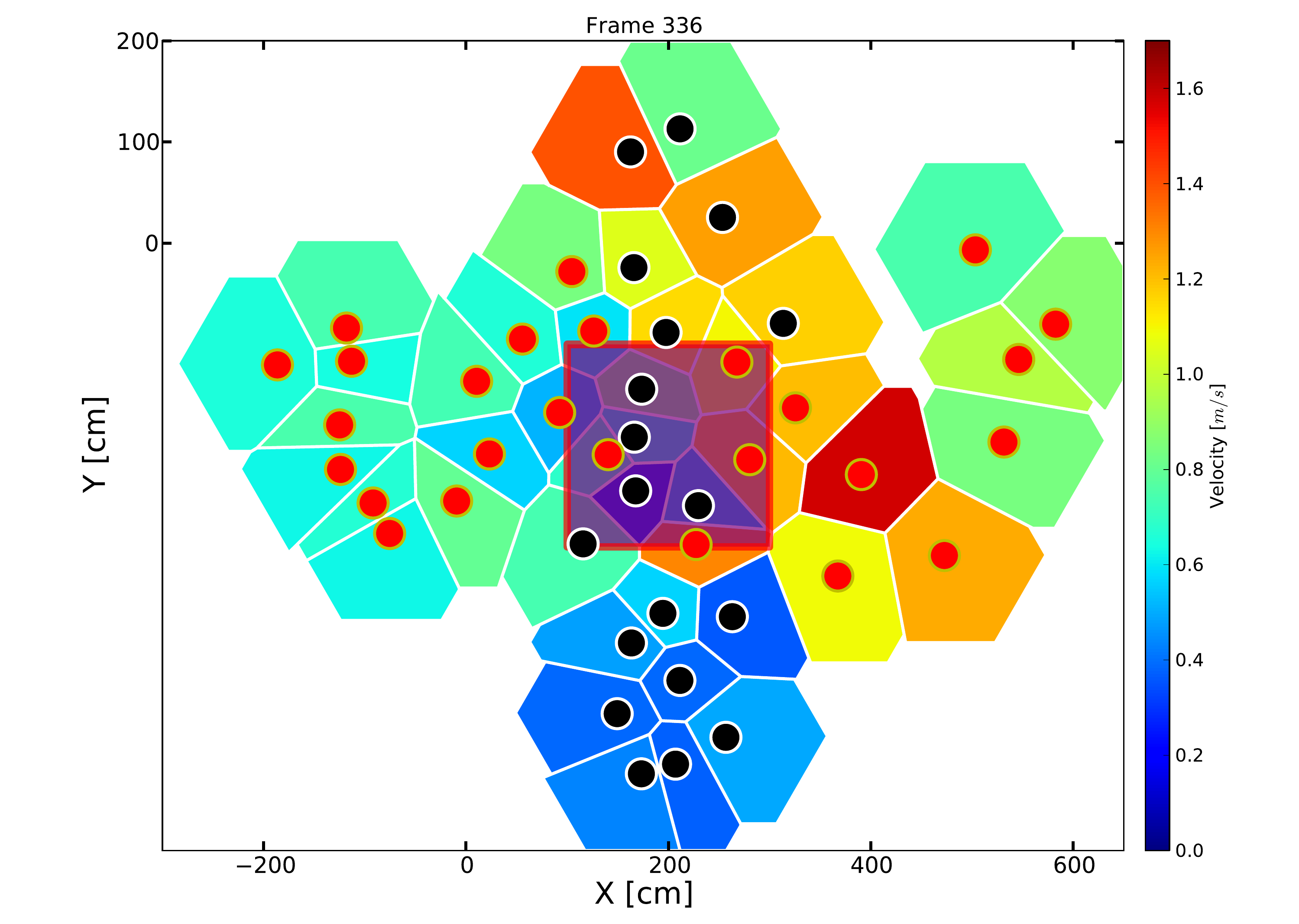}}
\caption{\label{fig-voro-cell} Voronoi method based density and velocity at time $t$. In Scenario 2 to 4, each Voronoi is cut by a hexagon with side length of $1~m$. The pedestrians moving to different directions are represented by different colors. The red squares in the middle shows the measurement areas of the Voronoi density and velocity. }
\end{figure}

In Scenario 2 to 4 the pedestrian streams use only a part of the available space, whereas the whole space between the walls are used in Scenario 1. To compare the densities, flows and velocities from different geometries and systems, we aim for reducing the effects of the boundaries and the measurement area. For the former, each pedestrian are limited in a hexagon with the side length of $1~m$ and each Voronoi cell are cut by the hexagon for Scenario 2 to 4. In such way, the Voronoi cells of boundary pedestrians are no longer infinite and their influence on the measured results can be reduced (see Table~\ref{table-area}). To analyze the influence of measurement area, we select different sizes of measurement area whose center locates in the middle of stream to make stable analysis. However, no stable density and velocity can be measured from these less particle system. The results from large area have small fluctuation but strongly depend on the boundary effect. Nevertheless, small area can reduce the boundary effect but causes large fluctuations (see Figure~\ref{fig-voro-measurearea} and Table~\ref{table-area}). In this study, the moderate size $2~m \times 2~m$ are selected as measurement area (see Figure~\ref{fig-voro-cell}) for all systems to make comparison. With this selection, the influence of boundary effect is limited and the fluctuation of measured values is also not so large. we measure the fundamental diagram of the flows in the middle of the streams where the outermost layer of stream is not totally included.

\renewcommand{\baselinestretch}{1.2}\normalsize
\begin{table}[htbp]
 \begin{tabular}{cccccc}
  \toprule
  Measurement  & Scenario 1 & Scenario 2  & Scenario 2  & Scenario 4  & Scenario 4  \\
  area [$m^2$] &  &  (without cutting) &  (cutting) & (without cutting) & (cutting)  \\
  \midrule
  $1.0 \times 1.0$ & $ 1.68 \pm 0.35 $ & $ 1.45 \pm 0.30 $ & $ 1.48 \pm 0.29 $ & $ 2.06 \pm 0.26 $ & $ 2.06 \pm 0.26$\\
  $1.5 \times 1.5$ & $ 1.64 \pm 0.29 $ & $ 1.39 \pm 0.20 $ & $ 1.43 \pm 0.20 $ & $ 1.90 \pm 0.18 $ & $ 1.90 \pm 0.18$\\
  $1.7 \times 1.7$ & $ 1.62 \pm 0.28 $ & $ 1.37 \pm 0.18 $ & $ 1.42 \pm 0.18 $ & $ 1.84 \pm 0.16 $ & $ 1.84 \pm 0.15$\\
  $2.0 \times 2.0$ & $ 1.60 \pm 0.27 $ & $ 1.31 \pm 0.16 $ & $ 1.38 \pm 0.16 $ & $ 1.75 \pm 0.12 $ & $ 1.76 \pm 0.12$\\
  $2.3 \times 2.3$ & $ 1.56 \pm 0.25 $ & $ 1.27 \pm 0.14 $ & $ 1.35 \pm 0.14 $ & $ 1.67 \pm 0.10 $ & $ 1.68 \pm 0.10$\\
  $2.5 \times 2.5$ & $ 1.54 \pm 0.24 $ & $ 1.24 \pm 0.13 $ & $ 1.33 \pm 0.14 $ & $ 1.61 \pm 0.09 $ & $ 1.63 \pm 0.09$\\
  $3.0 \times 3.0$ & $ 1.49 \pm 0.23 $ & $ 1.15 \pm 0.11 $ & $ 1.26 \pm 0.11 $ & $ 1.48 \pm 0.08 $ & $ 1.52 \pm 0.08$\\
  $3.6 \times 3.6$ & $ 1.47 \pm 0.21 $ & $ 1.00 \pm 0.09 $ & $ 1.11 \pm 0.10 $ & $ 1.33 \pm 0.07 $ & $ 1.40 \pm 0.06$\\
  \bottomrule
 \end{tabular}
 \centering\caption{\label{table-area} Data source of density ($m^{-2}$) in Figure~\ref{fig-voro-measurearea}. Here 'cutting' means that the Voronoi cells are cut by a hexagon with side length of $1~m$.}
\end{table}
\renewcommand{\baselinestretch}{1.2}\normalsize

\begin{figure}
\centering\subfigure[Scenario 2]{
\includegraphics[width=0.4\textwidth]{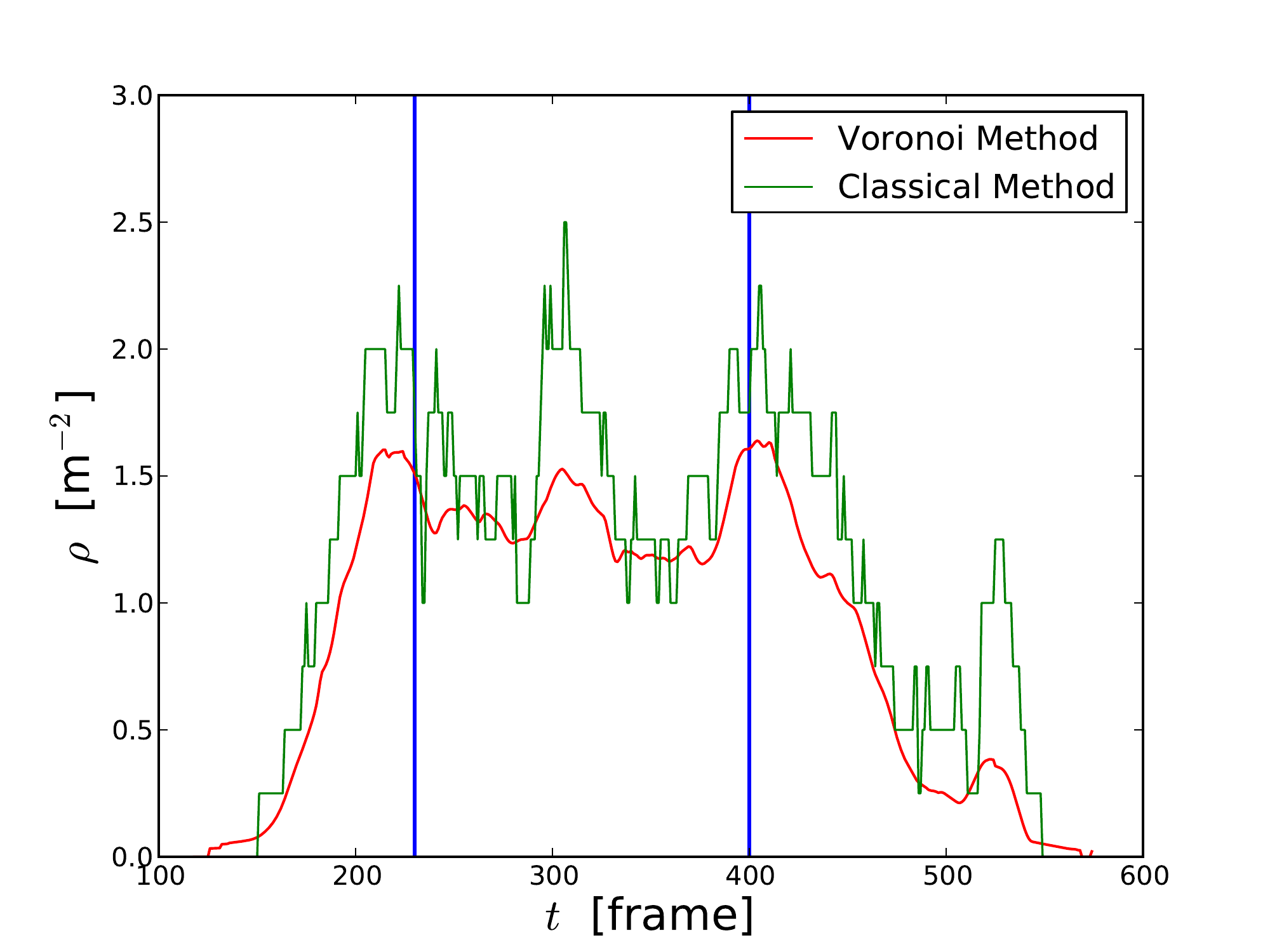}
\includegraphics[width=0.4\textwidth]{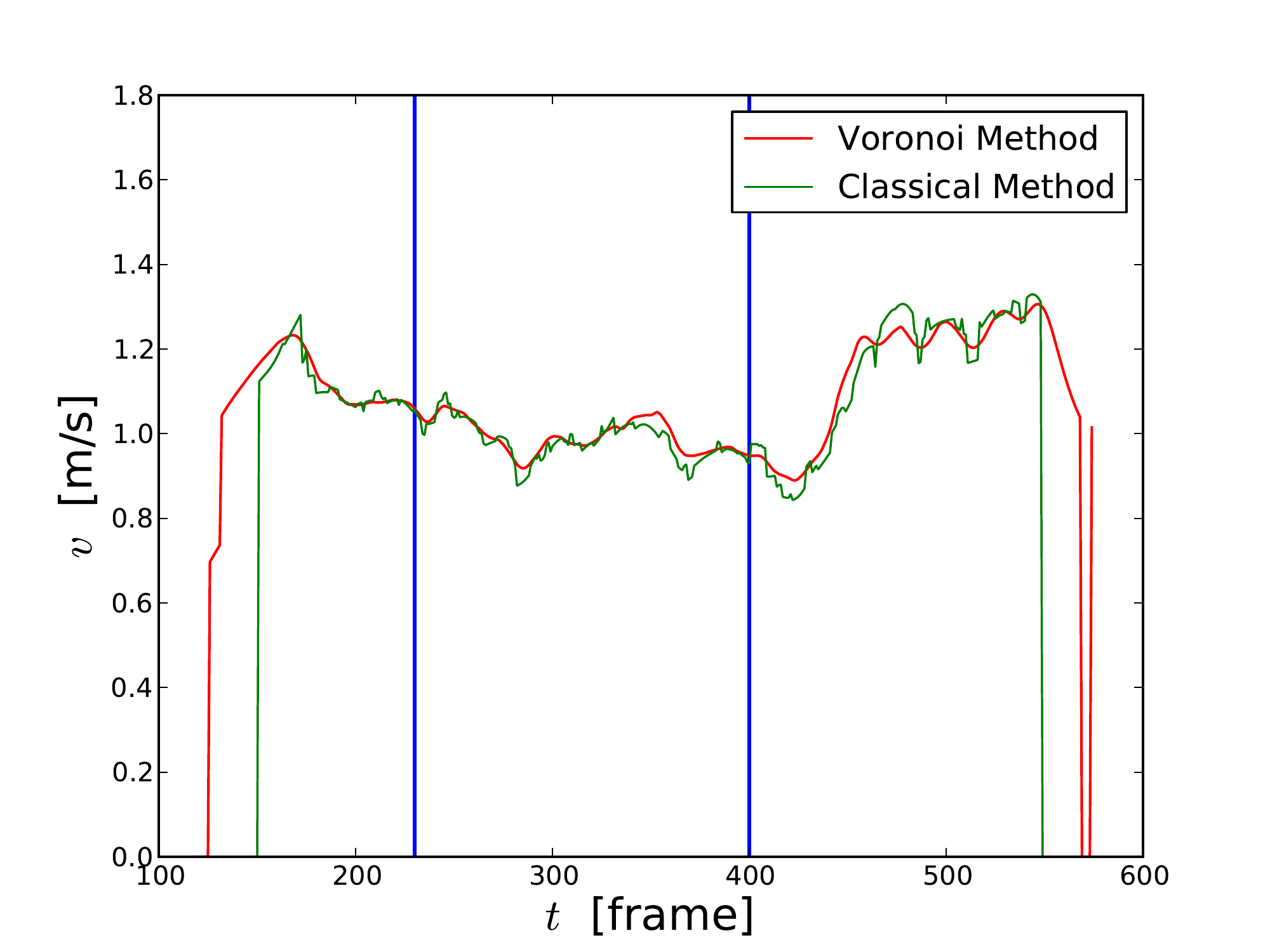}}
\centering\subfigure[Scenario 3]{
\includegraphics[width=0.4\textwidth]{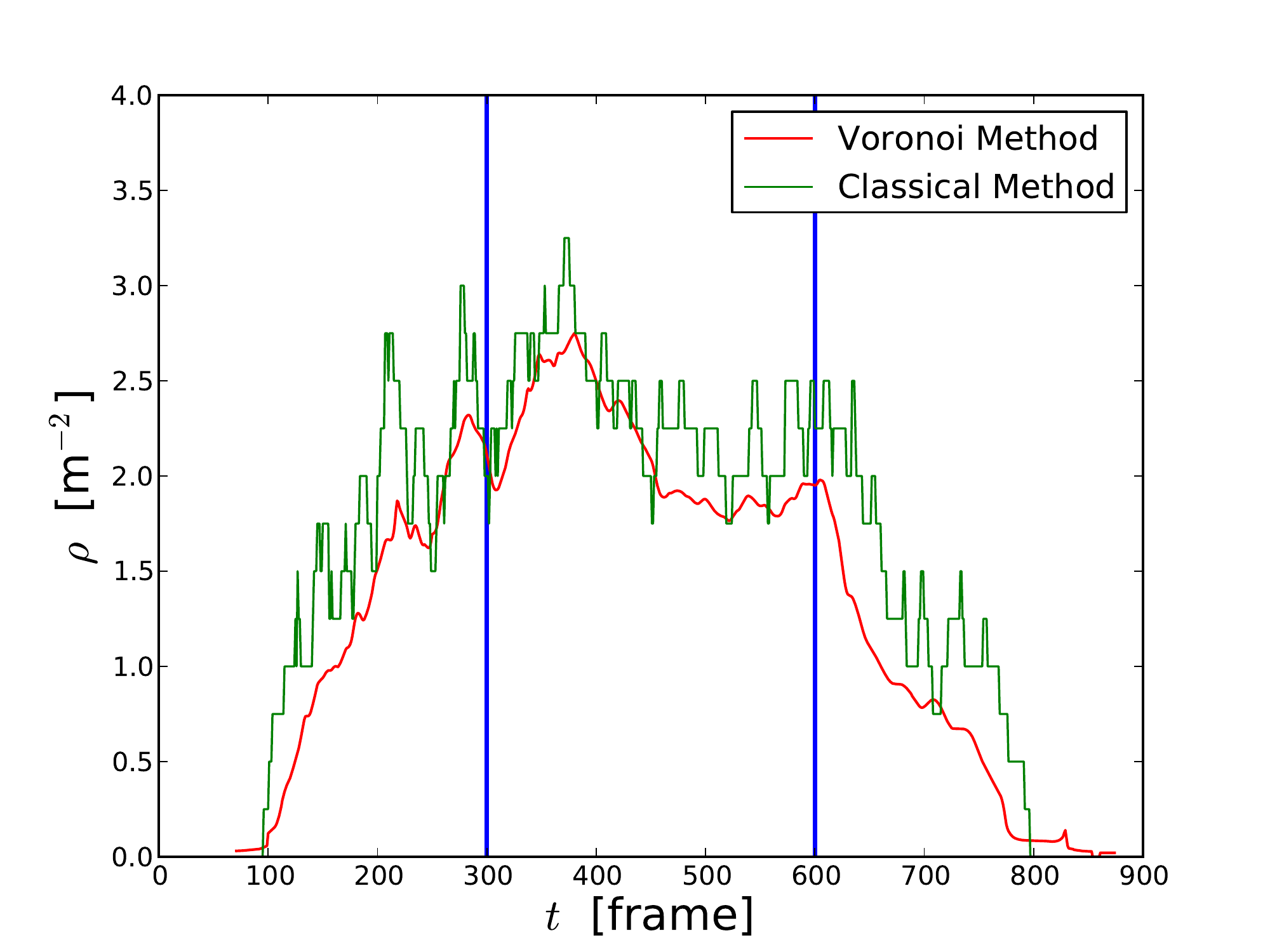}
\includegraphics[width=0.4\textwidth]{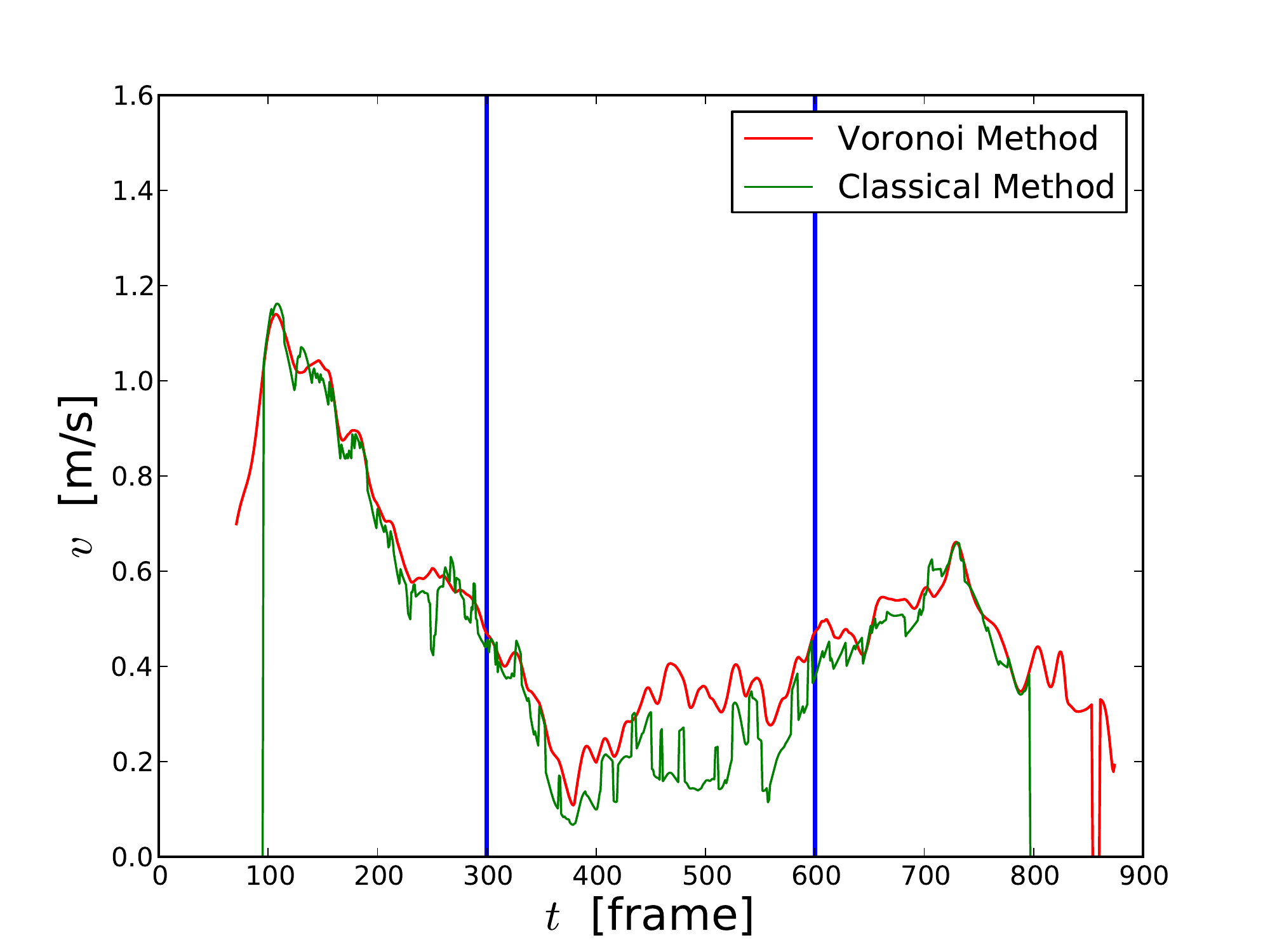}}
\centering\subfigure[Scenario 4]{
\includegraphics[width=0.4\textwidth]{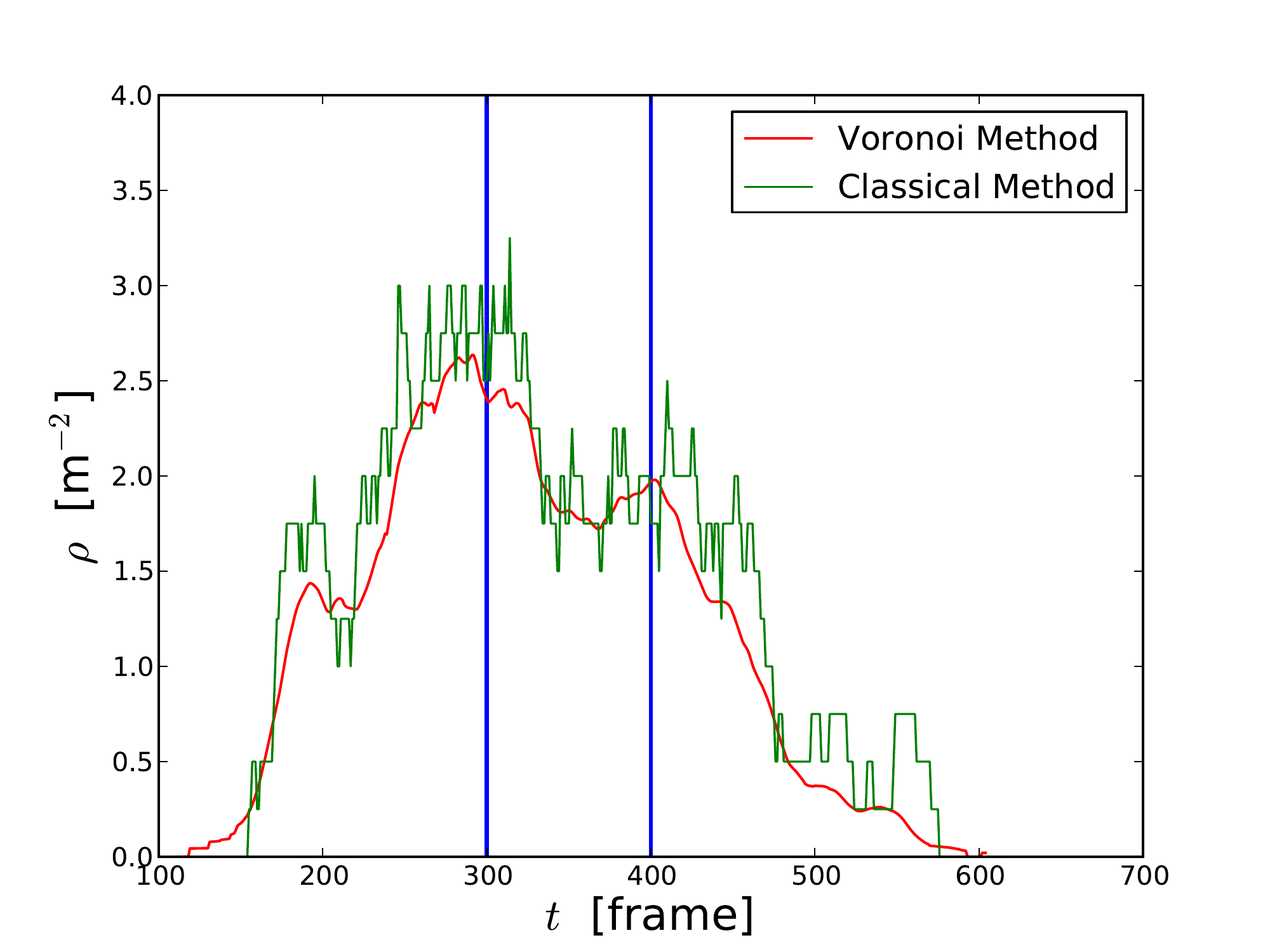}
\includegraphics[width=0.4\textwidth]{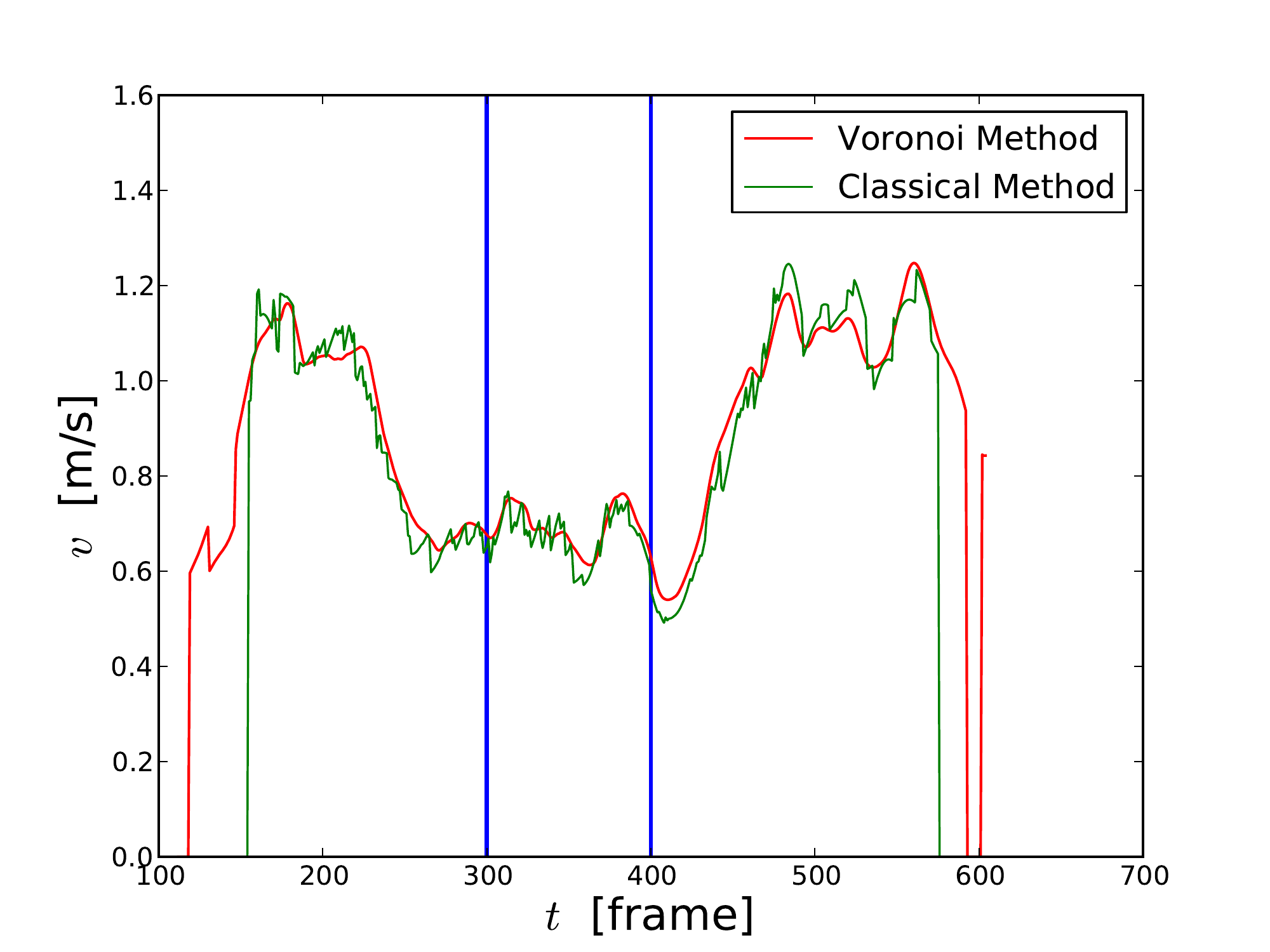}}
\caption{\label{fig-time-series} Time series of density and velocity. The two vertical lines in each diagram show the start and the end of stationary state, which was chosen manually in this study.}
\end{figure}

Figure~\ref{fig-time-series} shows the time series of mean density and velocity over the measurement area of the runs for last three scenarios. The density increases to certain value with time and keeps constant for some time. At the later stage of each run,  the number of pedestrians entering the measurement area is less and less, which lead to the decrease of mean density and velocity. In this study we try to investigate the steady but not transient behavior of the system. Thus we only use the period when the density and velocity remain relatively stable in the time series. And the other data points at the beginning and the end of the run were discarded in the following analysis.

%
%

\subsection{Crossing Behavior on Fundamental Diagram}

\begin{figure}
\centering{
\includegraphics[width=0.45\textwidth]{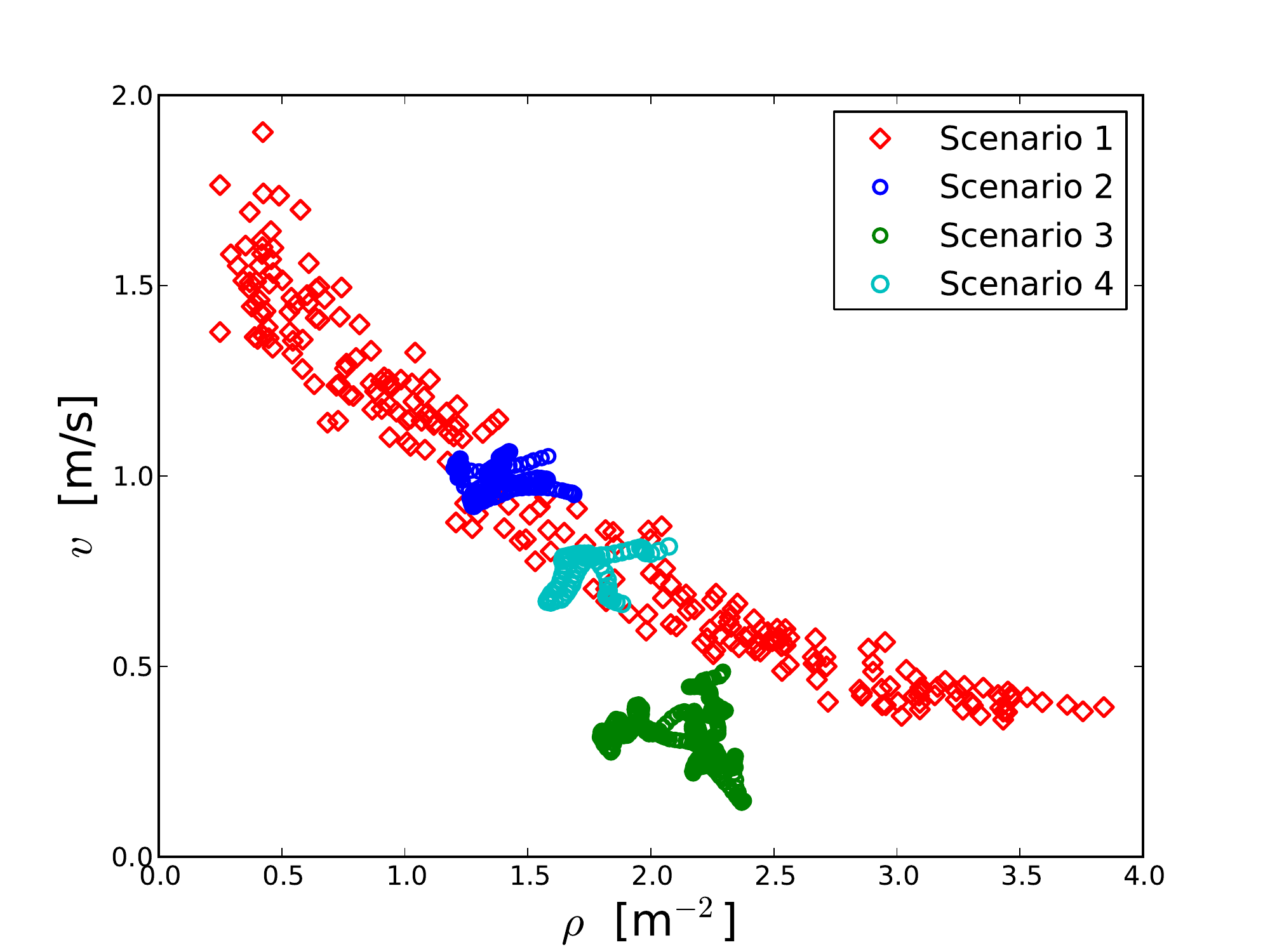}
\includegraphics[width=0.45\textwidth]{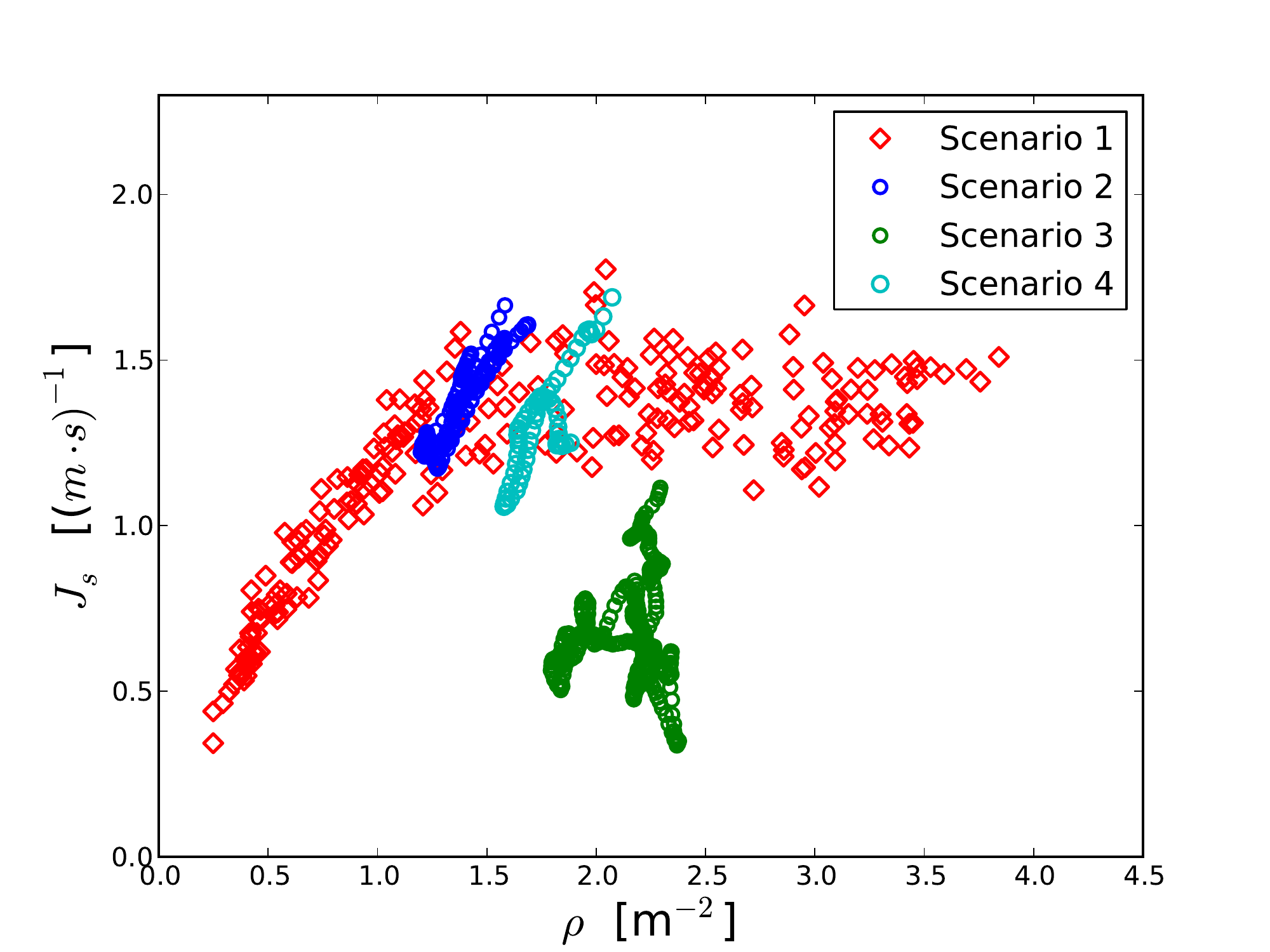}}
\caption{\label{fig-ped-FD} Comparison of fundamental diagrams from different scenarios.}
\end{figure}

In this section, we try to study the fundamental diagram and the effect of crossing behavior on it.
Compared to experiments in Hermes project, pedestrians at the outermost layer of the streams have relatively larger personal space and their movements are much freer. Reflected in Figure~\ref{fig-voro-cell}, it can be seen that the density and velocity are not homogenous over space. The pedestrians at outermost boundary have higher velocities and larger personal spaces. To study the influence of boundary condition, we compare the fundamental diagram of Scenario 1 and 2. Both of them were built on plane ground and in the form of bidirectional flow with DML. The number of lanes in the corridor changes from time to time and the maximum are 4 in Scenario 1 and 3 in Scenario 2. Head-on and cross-directional conflicts in this kind of flow mainly occur at the interface of the lanes. Without boundary pedestrians in Scenario 2 could be easier to keep a certain distance between lanes to avoid conflicts. However, the maximal density in this run is not larger than  $1.5~m^{-2}$, which belongs to free flow state. According to the findings in \cite{Zhang2012}, no difference can be observed for different types of bidirectional flow  in straight corridor for $\rho < 2.0~m^{-2}$. This point is again proved from the comparison between the results from Scenario 1 and 2 in Figure~\ref{fig-ped-FD}.

A part of the stairway and the two pillars on the plane ground influence the stream in Scenario 3. On one hand, the available width of the stairway is smaller than the potential width of the plane corridor. On the other hand, the characteristics of pedestrian movement on staircase, no matter upward or downward, is different from plane ground. At the same density, the speed and specific flow are relatively smaller on stairs \cite{Burghardt2013}. Consequently, the transition region from the plane ground to the stair acts like a bottleneck, where a tailback occurs influencing the dynamic at the crossing. On the plane ground the stream to the bottleneck was influenced by the congestion, while the other stream away from the bottleneck was less influenced by it. Reflected in Figure~\ref{fig-voro-cell}(b), the velocity distribution over space shows no homogeneous and the velocities for the pedestrians who do not pass the pillar with the main stream in the center but in the right side are obviously higher. Comparing the fundamental diagram from Scenario 1 and 3, we could find that not only velocities but also specific flow are smaller at the same density when a bottleneck is introduced in the scenario of bidirectional flow.


In Scenario 4, bidirectional flow with a $90^\circ$ intersecting angle was achieved. The conflicts among pedestrians are not head-on conflicts any more but from the side. In such situation, it is observed that pedestrians try to cross the flow obliquely rather than vertically \cite{Cividini2013}. This self-organized coordination behavior  reduces conflicts largely and therefore makes the flow sharper and smooth. In this run, the stairway was still used by one stream, in which the pedestrians move up stairs to the plane ground. At about $t = 250~frame$, the surfaces of the two streams began to contact, the velocity was affected and dropped rapidly from $1.1~m/s$ to $0.7~m/s$. From the video recordings, we can see that two streams both exist in the measurement area from $t = 300~frame$.  Then pedestrians take some time to regulate the crossing behavior persistently to a stationary state. After $t = 400~frame$ only one stream exist in the system. Thus, we collected the data points from $t = 300$ to $400~frame$ to compare with that from Scenario 1 in Figure~\ref{fig-ped-FD}. We can see that they are consistent with each other in the small density range. In crossing stream, pedestrians try to coordinate their movement and decrease potential conflicts. When the flow is at stationary state, no apparent difference can be observed compared to other kind of bidirectional flow.


\section{Summary}\label{sec-summary}

Series of controlled laboratory pedestrian experiments of bidirectional flow were compared. The experiments were recorded using video cameras and the trajectories of all pedestrian were extracted with high accuracy. To gain high resolution in time and space in combination with small fluctuation the Voronoi method is used to determine the fundamental diagram. The influences of boundary condition on the velocity and density are studied. No apparent differences are observed except that the transition to the stair in Scenario 3 acts like a bottleneck and leads to lower flow values. For Scenario 1 we were able to determine the fundamental diagram of bidirectional flow with an intersecting angle of $180^\circ$ from low densities up to a density of $3.8~m^{-2}$. In Scenario 2 the intersecting angle was smaler than $180^\circ$ and the resulting velocities correspond to the velocities of Scenario 1. Even in Scenario 4, where the intersecting angle of the flow was $90^\circ$, the results show no differences to the other fundamental diagrams.

The data compared in this study indicate that intersecting angle has no influence on the fundamental diagram. Thus the processes responsible for the reduction of the velocity or flow in comaprison to unidirectional streams, like stearing maneuvers to avoid collisions, have the same nature and strength is and are independent from the intersecting angle.

A lot of questions regarding the transport properties of intersecting streams of self-driven particles like pedestrians are open. In the future we plan to systematically study the influence of the number and angel of intersecting streams.

%
%
%
%

\section*{References}
\bibliographystyle{IEEEtran}
\bibliography{ped}

\begin{thebibliography}{10}
\providecommand{\url}[1]{#1}
\csname url@samestyle\endcsname
\providecommand{\newblock}{\relax}
\providecommand{\bibinfo}[2]{#2}
\providecommand{\BIBentrySTDinterwordspacing}{\spaceskip=0pt\relax}
\providecommand{\BIBentryALTinterwordstretchfactor}{4}
\providecommand{\BIBentryALTinterwordspacing}{\spaceskip=\fontdimen2\font plus
\BIBentryALTinterwordstretchfactor\fontdimen3\font minus
  \fontdimen4\font\relax}
\providecommand{\BIBforeignlanguage}[2]{{%
\expandafter\ifx\csname l@#1\endcsname\relax
\typeout{** WARNING: IEEEtran.bst: No hyphenation pattern has been}%
\typeout{** loaded for the language `#1'. Using the pattern for}%
\typeout{** the default language instead.}%
\else
\language=\csname l@#1\endcsname
\fi
#2}}
\providecommand{\BIBdecl}{\relax}
\BIBdecl

\bibitem{duisburg}
``http://en.wikipedia.org/wiki/love-parade-stampede.''

\bibitem{Zhang2012}
\BIBentryALTinterwordspacing
J.~Zhang, W.~Klingsch, A.~Schadschneider, and A.~Seyfried, ``Ordering in
  bidirectional pedestrian flows and its influence on the fundamental
  diagram,'' \emph{Journal of Statistical Mechanics: Theory and Experiment},
  vol. 2012, no.~02, p. P02002, 2012. 
\BIBentrySTDinterwordspacing

\bibitem{HCM2000}
T.~R. Board, ``Highway capacity manual,'' Transportation Research Board,
  Washington DC, Tech. Rep., 2000.

\bibitem{Lam2002}
W.~H.~K. Lam, J.~Y.~S. Lee, and C.~Y. Cheung, ``A study of the bi-directional
  pedestrian flow characteristics at hong kong signalized crosswalk
  facilities,'' \emph{Transportation}, vol.~29, pp. 169--192, 2002.

\bibitem{Lam2003}
W.~H.~K. Lam, J.~Y.~S. Lee, K.~S. Chan, and P.~K. Goh, ``A generalised function
  for modeling bi-directional flow effects on indoor walkways in hong kong,''
  \emph{Transportation Research Part A: Policy and Practice}, vol.~37, pp.
  789--810, 2003.

\bibitem{Alhajyaseen2011}
\BIBentryALTinterwordspacing
W.~K. Alhajyaseen, H.~Nakamura, and M.~Asano, ``Effects of bi-directional
  pedestrian flow characteristics upon the capacity of signalized crosswalks,''
  \emph{Procedia - Social and Behavioral Sciences}, vol.~16, pp. 526 -- 535,
  2011, 6th International Symposium on Highway Capacity and Quality of Service.
\BIBentrySTDinterwordspacing

\bibitem{Navin1969}
F.~D. Navin and R.~J. Wheeler, ``Pedestrian flow characteristics,''
  \emph{Traffic Engineering}, vol.~39, pp. 30--36, 1969.

\bibitem{O'Flaherty1972}
C.~A. O'Flaherty and M.~H. Parkinson, ``Movement on a city centre footway,''
  \emph{Traffic Engineering and Control}, vol.~13, pp. 434--438, Feb. 1972.

\bibitem{Polus1983}
A.~Polus, J.~L. Schofer, and A.~Ushpiz, ``Pedestrian flow and level of
  service,'' \emph{Journal of Transportation Engineering}, vol. 109 1, pp.
  46--56, 1983.

\bibitem{Tanaboriboon1986}
Y.~Tanaboriboon, S.~S. Hwa, and C.~H. Chor, ``Pedestrian characteristics study
  in singapore,'' \emph{Journal of Transportation Engineering}, vol. 112, pp.
  229--235, 1986.

\bibitem{Jian2010}
M.~Jian, S.~Weiguo, Z.~Jun, L.~Siuming, and L.~Guahgxuan, ``k-nearest-neighbor
  interaction induced self-organized pedestrian counter flow,'' \emph{Physica
  A: Statistical Mechanics and its Applications}, vol. 389, no.~10, pp.
  2101--2117, may 2010.

\bibitem{Older1968}
S.~Older, ``Movement of pedestrians on footways in shopping streets,''
  \emph{Traffic Engineering and Control}, vol.~10, pp. 160--163, 1968.

\bibitem{Hughes2000}
R.~L. Hughes, ``The flow of large crowds of pedestrians,'' \emph{Mathematics
  and Computer in Simulation}, vol.~53, pp. 367--370, 2000.

\bibitem{Hoogendoorn2003153}
S.~Hoogendoorn and P.~Bovy, ``Simulation of pedestrian flows by optimal control
  and differential games,'' \emph{Optimal Control Applications and Methods},
  vol.~24, no.~3, pp. 153--172, 2003, cited By (since 1996)53.

\bibitem{Cividini2013}
J.~Cividini, C.~Appert-Rolland, and H.~J. Hilhorst, ``Diagonal patterns and
  chevron effect in intersecting traffic flows,'' \emph{Europhysics Letters},
  vol. 102, p. 20002, 2013.

\bibitem{Wong2010}
\BIBentryALTinterwordspacing
S.~C. Wong, W.~L. Leung, S.~H. Chan, W.~H.~K. Lam, N.~H. Yung, C.~Y. Liu, and
  P.~Zhang, ``Bidirectional pedestrian stream model with oblique intersecting
  angle,'' \emph{Journal of transportation engineering}, vol. 136, no.~3, pp.
  234--242, 2010.
\BIBentrySTDinterwordspacing

\bibitem{Guo2012c}
G.~Ren-Yong, W.~S. C., X.~Yin-Hua, H.~Hai-Jun, L.~W.~H. K., and C.~Keechoo,
  ``Empirical evidence for the look-ahead behavior of pedestrians in
  bi-directional flows,'' \emph{CHIN. PHYS. LETT}, vol.~29, pp. 2 -- 5, March
  2012.

\bibitem{Holl2009}
\BIBentryALTinterwordspacing
S.~Holl and A.~Seyfried, ``Hermes - an evacuation assistant for mass events,''
  \emph{inSiDe}, vol.~7, no.~1, pp. 60--61, 2009. [Online]. Available:
\BIBentrySTDinterwordspacing

\bibitem{KemlohWagoum201398}
A.~U.~K. Wagoum, B.~Steffen, A.~Seyfried, and M.~Chraibi, ``Parallel real time
  computation of large scale pedestrian evacuations,'' \emph{Advances in
  Engineering Software}, vol. 60-61, no.~0, pp. 98 -- 103, 2013.

\bibitem{Plaue2011}
\BIBentryALTinterwordspacing
M.~Plaue, M.~Chen, G.~B\"{a}rwolff, and H.~Schwandt, ``Trajectory extraction
  and density analysis of intersecting pedestrian flows from video
  recordings,'' in \emph{Proceedings of the 2011 ISPRS conference on
  Photogrammetric image analysis}, ser. PIA'11.\hskip 1em plus 0.5em minus
  0.4em\relax Berlin, Heidelberg: Springer-Verlag, 2011, pp. 285--296.
\BIBentrySTDinterwordspacing

\bibitem{Zhang2011}
J.~Zhang, W.~Klingsch, A.~Schadschneider, and A.~Seyfried, ``Transitions in
  pedestrian fundamental diagrams of straight corridors and t-junctions,''
  \emph{Journal of Statistical Mechanics: Theory and Experiment}, vol. P06004,
  jun 2011.

\bibitem{Burghardt2013}
\BIBentryALTinterwordspacing
S.~Burghardt, A.~Seyfried, and W.~Klingsch, ``Performance of stairs
  鈥�fundamental diagram and topographical measurements,''
  \emph{Transportation Research Part C: Emerging Technologies}, no.~0, pp.~--,
  2013.
\BIBentrySTDinterwordspacing

\end{thebibliography}
\end{document}